\documentclass[10pt,journal]{IEEEtran}
\usepackage{amsmath,amssymb,amsfonts}
\usepackage{graphicx}
\usepackage{multirow}
\usepackage{cuted}
\usepackage{CJK}
\usepackage{indentfirst}
\usepackage{xcolor}
\usepackage{hyperref}
\usepackage{epstopdf}
\usepackage{bm}
\usepackage{cases}
\usepackage{subfigure}
\usepackage{caption}
\usepackage{bookmark}
\usepackage{makecell}

\usepackage{algorithmic}
\usepackage{algorithm}
\usepackage{array}
\usepackage{textcomp}
\usepackage{stfloats}
\usepackage{url}
\usepackage{verbatim}
\usepackage{cite}
\hypersetup{
colorlinks=true,
linkcolor=black,
citecolor=black
}
\begin{document}

\title{Discrete-Time CRLB-based Power Allocation for CF MIMO-ISAC with Joint Localization and Velocity Sensing}
\author{Guoqing Xia, Qu Luo, \textit{Member, IEEE}, Pei Xiao, \textit{Senior Member, IEEE}, Bing Ji, \textit{Senior Member, IEEE}, Yue Zhang, \textit{Senior Member, IEEE}, Cicek Cavdar, \textit{Member, IEEE} and Huiyu Zhou.

\thanks{Guoqing Xia and Bing Ji are with the School of Engineering, University of Leicester, LE1 7RH
Leicester, UK (e-mail: \{gx21, bing.ji\}@leicester.ac.uk). 
Pei Xiao and Qu Luo are with 5GIC $\&$ 6GIC, Institute for Communication Systems (ICS) of University of Surrey, Guildford, GU2 7XH, UK (e-mail: \{p.xiao, q.u.luo\}@surrey.ac.uk). 
Yue Zhang is with the Institute for Communication Systems and Measurement of China, Chengdu, 610095, China (e-mail: zhangyue@icsmcn.cn). 
Cicek Cavdar is with the School of Electrical Engineering and Computer Science, KTH Royal Institute of Technology, Brinellvägen 8, 114 28, Stockholm, Sweden (email: cavdar@kth.se). Huiyu Zhou is with the School of Computing and Mathematical Sciences, University of Leicester, Leicester, LE1 7RH, UK (e-mail: hz143@leicester.ac.uk).}}

\maketitle
\begin{abstract}
This paper investigates integrated sensing and communication (ISAC) in a cell-free (CF) multiple-input multiple-output (MIMO) network, in which each access point operates either as an ISAC transmitter or as a sensing receiver. Discrete-time signal-based Cramér–Rao lower bounds (CRLBs) are employed as sensing performance metrics for joint location and velocity estimation under arbitrary power allocation ratios, assuming a deterministic radar cross section (RCS) model. Based on these metrics, a power allocation problem for CF MIMO-ISAC is formulated to maximize the communication signal-to-interference-plus-noise ratio (SINR), subject to CRLB-based sensing constraints and per-transmitter power limits.
To address the resulting nonlinear and nonconvex optimization problem, a penalty-function and projection-based modified conjugate gradient algorithm with inexact line search (PP-MCG-ILS) is developed, together with an alternative method based on modified steepest descent (PP-MSD-ILS). The proposed algorithms are applicable to a class of constrained optimization problems involving nonlinear inequality constraints and affine equality constraints. In addition, the PP-MCG-ILS algorithm is extended to a sensing-only scenario, where a penalty-function-based normalized conjugate gradient algorithm (P-NCG-ILS) is employed for sensing power minimization.
The convergence behavior of the proposed algorithms is analyzed, and their computational complexity is qualitatively evaluated. Simulation results validate the accuracy of the derived CRLBs and demonstrate the effectiveness of the proposed power allocation strategies in enhancing sensing performance and overall ISAC performance.
\end{abstract}

\begin{IEEEkeywords}
   Cell-free MIMO ISAC, CRLB, Power allocation, Penalty function, Projection operator, Conjugate gradient.
\end{IEEEkeywords}

\vspace{-0.5em}
\section{Introduction}
\IEEEPARstart{I}{ntegrated} sensing and communication (ISAC) has emerged as a key enabling technology for sixth-generation (6G) applications, including smart cities, autonomous driving, and remote healthcare \cite{Liu2022}. Unlike traditional systems that separate the design of communication and sensing functionalities, ISAC integrates both within a unified framework using shared hardware platforms and spectral resources. This integration offers substantial improvements in spectral efficiency, energy savings, and hardware utilization \cite{Liu2022survey, Xiong2023}.


ISAC systems can be broadly categorized based on the spatial configuration of transmitters and receivers. Monostatic ISAC systems, where the transmitter and receiver are co-located, typically rely on advanced beamforming techniques to mitigate self-interference in full-duplex operations \cite{Liu2018, Barneto2021}. In contrast, multistatic ISAC systems deploy spatially separated transmitters and receivers, facilitating spatial diversity and sensing coverage \cite{Dehkordi2024,Sruti2025}. Building upon the multistatic architecture, the concept of cell-free (CF) multiple-input and multiple-output (MIMO) ISAC has recently emerged \cite{Femenias2025,Xu2025}. In CF MIMO-ISAC systems, a large number of distributed access points (APs) jointly serve communication users and provide sensing services in a coordinated manner without relying on conventional cell boundaries. This architecture inherits the advantages of both distributed multistatic radar and CF MIMO communication, offering improved spectral efficiency, spatial resolution, and system scalability \cite{Nuria2024}.


Recent advancements in CF MIMO-ISAC have extended from basic trade-off analysis \cite{Weihao2024} to complex sensing tasks, including target detection \cite{An2023}, and parameter estimation for radar cross-section (RCS) \cite{Qi2022}, location \cite{Zhang2025}, velocity \cite{Wang2025}, and angle \cite{Naoumi2024}. For example, \cite{An2023} studies the fundamental performance tradeoff between the \emph{target detection probability} and the user’s achievable rate in ISAC systems by deriving the probability of false alarm and the successful probability of detection for the target of interest. \cite{Qi2022} provides a unified framework integrating sensing (\emph{RCS estimation}), computing, and communication to optimize limited system resource. \cite{Zhang2025} proposes a two-stage scheme for \emph{target localization}, with communication signals reused as sensing reference signals. Additionally, joint sensing and communication beamforming has been explored in \cite{Demirhan2024}.

Despite these advances, a fundamental challenge in ISAC system design, including CF MIMO-ISAC,  is the formulation of robust performance metrics, which are critical for guiding system design and balancing communication-sensing trade-offs \cite{Liu2022survey, Xiong2023}. While unified metrics based on mutual information \cite{Ouyang2023MI} and Kullback–Leibler divergence \cite{Mohammad2023} have been proposed for general analysis, adapting these metrics to task-specific sensing objectives remains a significant challenge. Consequently, task-specific metrics for sensing and communication are still widely adopted.

For communication, Shannon capacity underpins metrics such as spectral efficiency \cite{Ammar2022Cell_free}, achievable rate \cite{Ngo2017}, and signal-to-interference-plus-noise ratio (SINR). Additional metrics, including coverage probability and energy efficiency, have also been considered, particularly in scenarios with stringent quality-of-service (QoS) or energy constraints \cite{Gan2024}. 
 On the sensing side, a variety of performance metrics have been investigated, such as sensing mutual information \cite{Sun2020MI}, sensing SINR \cite{Behdad2024, Fan2025}, and the Cramér–Rao lower bound (CRLB) \cite{He2010, He2016, Godrich2010, Ai2015}. Among these, CRLB-based formulations have proven particularly effective for parameter estimation and algorithm benchmarking. CRLB-based approaches generally fall into two categories depending on the RCS model: deterministic and stochastic.
Under the stochastic RCS assumption, RCS values are modeled as random variables, and the corresponding likelihood functions are derived by averaging over their statistical distributions.
In contrast, the deterministic RCS assumption treats RCS values as deterministic but unknown quantities that may vary spatially. This model is particularly applicable in slowly varying environments or when prior RCS knowledge is unavailable. CRLBs for both single-target and multiple-target location estimation have been analyzed under the stochastic RCS assumption \cite{He2010, He2016} and the deterministic RCS assumption \cite{Godrich2010, Ai2015}. 

However, most existing CRLB formulations are built upon continuous-time signal models, whereas practical radar systems operate in the discrete-time domain, where the sampling rate critically affects sensing accuracy. Our prior work \cite{xia2025cell} addresses this gap by introducing a discrete-time likelihood function over Gaussian channels, offering a more realistic modeling framework. Building on this foundation, this paper devotes into the CRLBs derivation for joint location and velocity estimation under the deterministic RCS assumption in CF MIMO-ISAC systems.

Beyond performance characterization, another critical challenge lies in resource allocation, involving time \cite{Zhang2024}, frequency \cite{Dong2023}, space \cite{Li2023}, and power \cite{Behdad2024, Fan2025, Ahmed2019,Huang2023, Ahmed2024}. Recent works in CF MIMO-ISAC systems have addressed power allocation (PA) under sensing SINR constraints \cite{Behdad2024, Fan2025} or sensing CRLB constraints \cite{Ahmed2019, Huang2023, Ahmed2024}. These involved sensing CRLBs typically consider localization only, neglecting velocity or angular estimation, and depend on continuous-time models that overlook the impact of sampling rate.

These considerations motivate us to consider a more comprehensive and realistic setting:  joint location and velocity estimation in CF MIMO-ISAC systems under a discrete-time signal model with a deterministic RCS assumption.  The  CRLBs are served as effective sensing metrics to guide PA across distributed APs, with the goal of maximizing communication SINR. 
This leads to a class of non-convex and highly nonlinear optimization problems.  To address this issue, we develop a tailored optimization framework that integrates penalty methods, projection operations, and gradient-based search, enabling efficient and reliable algorithmic solutions under complex practical constraints.
The main contributions of this work are summarized as follows:
\begin{itemize}
    \item We develop tractable closed-form approximations of the CRLB for joint location and velocity estimation under arbitrary power allocation (PA) strategies. The accuracy of the proposed approximations is validated through comparisons with maximum-likelihood estimation (MLE) results, indicating their suitability as sensing performance metrics for ISAC-oriented resource allocation. Based on these metrics, PA optimization problems are formulated to maximize communication SINR subject to sensing accuracy and power budget constraints.
    \item To address the nonconvex and nonlinear nature of the resulting PA optimization problems, we introduce a penalty-function and projection-based modified conjugate gradient algorithm with inexact line search (PP-MCG-ILS), along with a PP-MSD-ILS algorithm based on modified steepest descent. In addition, a penalty-function-based normalized conjugate gradient algorithm with inexact line search (P-NCG-ILS) is developed for power minimization in sensing-only scenarios under prescribed sensing accuracy constraints.
    \item We provide a detailed analysis of the convergence behavior and computational complexity of the proposed algorithms. Under standard smoothness assumptions, the analysis establishes well-defined step size selection, stable iterative updates, and efficient numerical behavior. The algorithmic framework also admits straightforward extension to related constrained nonconvex optimization problems.
    \item Simulation results are provided to evaluate the proposed algorithms in both ISAC and sensing-only scenarios. The results demonstrate that the proposed methods achieve reliable convergence and competitive PA performance relative to representative benchmark schemes, while revealing the inherent trade-offs between sensing accuracy and communication performance.
\end{itemize}

Section \ref{Sec:syst model} introduces the received signal model for the CF MIMO-ISAC system.
Section \ref{sec: Metric and Problem} formulates the PA problem for CF MIMO-ISAC based on the proposed metrics.
Section \ref{sec:algorithm derivation} presents the proposed PA algorithms for both CF MIMO-ISAC and the sensing-only scenario.
Section \ref{sec:Algorithm analysis} analyzes the convergence properties and computational complexity of the proposed algorithms.
Section \ref{sec:simulations} provides simulation results to validate the theoretical analysis and algorithmic performance.
Finally, Section \ref{sec:conclusion} concludes the paper.

\emph{Notation}: $\mathbb{R}$ and $\mathbb{C}$ denote the field of real and complex numbers, respectively. Scalars are denoted by lower-case letters, vectors and matrices respectively by lower- and upper-case boldface letters. The conjugate and transpose are denoted by $(\cdot)^*$ and $(\cdot)^{\rm T}$, respectively. $\mathbb{E}\{\cdot\}$ and $|\cdot|$ denote the mathematical expectation and element-wise modulus, respectively. $\|\cdot\|$ denotes the $l_2$ norm of a matrix. The notations $\min\{\cdot\}$ and $\max\{\cdot\}$ denote the minimum and maximum element of the enclosed set $\{\cdot\}$, respectively.

\vspace{-1em}
\section{Signal Model}\label{Sec:syst model}
\begin{figure}
   \centerline
   {\includegraphics[width=0.25\textwidth]{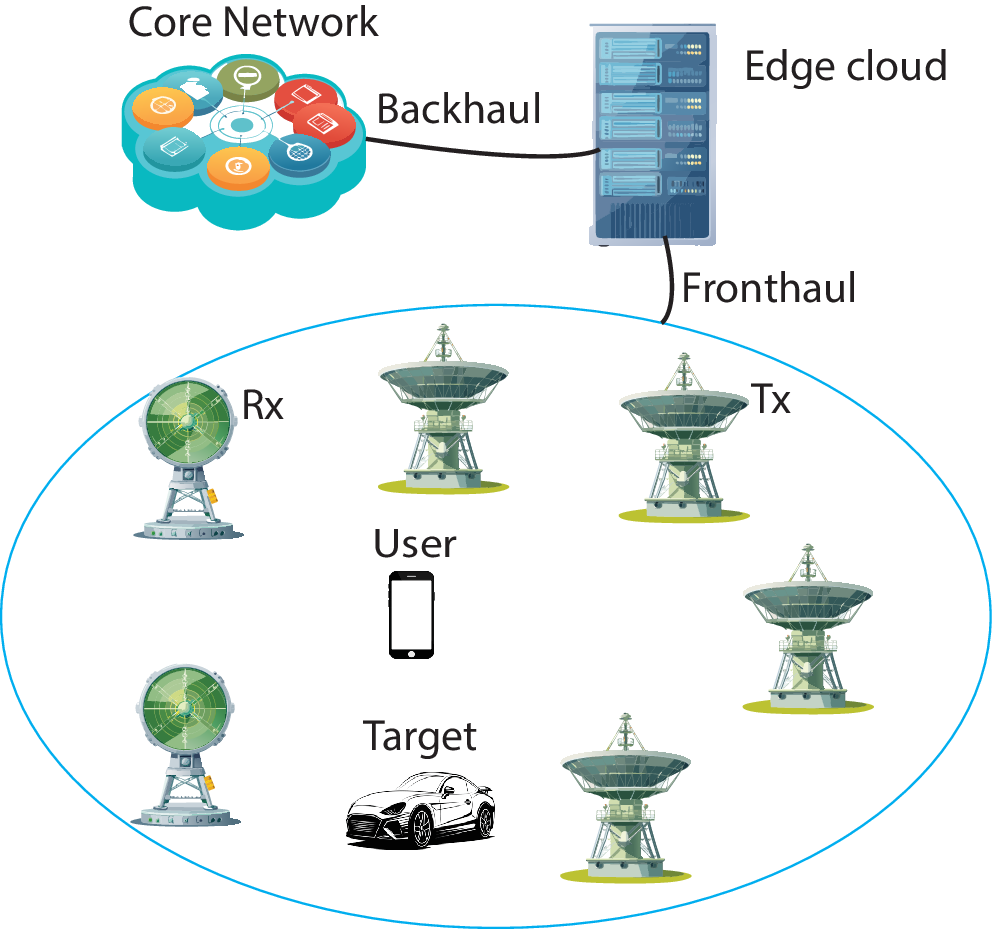}}
   \caption{CF MIMO system for ISAC.}\label{fig:MIMO_syst}
   \vspace{-1em}
\end{figure}

As shown in Fig. \ref{fig:MIMO_syst}, we consider a CF MIMO-ISAC system with widely separated APs. Each AP either serves as an ISAC transmitter or a sensing receiver, similar to the setup in \cite{Behdad2024}.  Assume that there are $N$ single-antenna  transmitters and $K$ single-antenna receivers, with the positions of the transmitter $n$ and receiver $k$ denoted as ${\bm l}_n = [x_n, y_n]^{\rm T}$ and ${\bm l}_k = [x_k, y_k]^{\rm T}$, respectively. All transmitters and receivers are scheduled and managed by the edge cloud and the core network, e.g., the fifth generation (5G) communication networks. Denote the location and velocity vectors of the sensing target as ${\bm l} = [x, y]^{\rm T}$ and ${\bm v} = [v_{x}, v_{y}]^{\rm T}$, respectively.

Denote $\tau_{n,k}$ as the propagation delay between the $n$th transmitter and the $k$th receiver, as reflected by the target. Then, we have
\begin{subequations}
   \begin{align}
\tau_{n,k}&=\tau_{n}+\tau_{k},\label{eq:trans delay}\\
      \tau_{n} ={\|{\bm l}_n-{\bm l}\|_2}/{c}, &  \quad    
      \tau_{k} ={\|{\bm l}_k-{\bm l}\|_2}/{c},\label{eq:rx delay}
   \end{align}
\end{subequations}   
where $\tau_{n}$ and $\tau_{k}$ represent the propagation delays from the $n$th transmitter to the $q$th target, and from the target to the $k$th receiver, respectively, and $c$ denotes the speed of light. The Doppler frequency induced by the relative radial velocity of the target with respect to transmitter $n$ and receiver $k$ is 
\begin{equation}
   f_{n,k} =  \dfrac{1}{\lambda}{\bm v}^{\rm T}\left(\frac{{\bm l}_n-{\bm l}}{\|{\bm l}_n-{\bm l}\|_2}+\frac{{\bm l}_k-{\bm l}}{\|{\bm l}_k-{\bm l}\|_2}\right), \label{eq:Doppler}
\end{equation}
where $\lambda$ denotes the wavelength.
 

A waveform $s_n(t)$,  with unit energy, i.e., $\varepsilon_{s}=\int |{s}_{n}(t)|^2\,dt=1$,  is considered for both sensing and communication at the $n$th transmitter.  Then, the baseband ISAC transmitted signal at transmitter $n$ is denoted as 
\begin{equation}
   {\bm x}_{n}(t) = \sqrt{P\rho_{n}}{b}_{n}\varsigma s_n(t), \label{eq:transmit sig sensing}
\end{equation}
where $P$ denotes the total available energy, $\rho_{n}$ is the PA factor with $\sum_{n=1}^N\rho_n=1$, ${b}_{n}\in \mathbb{C}$ is the normalized precoding weight with $|{b}_{n}|^2=1$, and $\varsigma\in \mathbb{C}$ has unit variance $\mathbb{E} \{|\varsigma|^2\}=1$. Let ${\bm \rho}=[\rho_1, \rho_2, \cdots, \rho_N]^{\rm T}$ be the PA coefficient vector over the $N$ transmitters. 

\textbf{Sensing model: } For target-reflected reception, we hereafter refer to the term ``RCS'' as encompassing both channel fading and radar RCS. Denote the RCS  associated with transmitter $n$ and receiver $k$ as ${\alpha}_{n,k}$.  The sensing model under the \emph{deterministic RCS assumption} is considered where RCSs are assumed to be deterministic but unknown during the observation time \cite{Ai2015,Godrich2010,van2001detection}.
Then, the lowpass equivalent of the reflected signal from the $n$th transmitter by a target received at receiver $k$ is given by\footnote{The transmitted signals over different transmitters are orthogonal and maintain approximate orthogonality over different delays and Doppler frequencies such that the signals contributed from different transmitters can be separated at each receiver \cite{Ai2015,He2010}.}
\begin{equation}
\begin{aligned}
   r_{n,k}(t) = \alpha_{n,k}y_{n,k}(t)+z_{n,k}(t),\label{eq:received reflected sig}  
\end{aligned}
\end{equation}
where $z_{n,k}(t)$ denotes the clutter-plus-noise and $y_{n,k}(t)$ is the loss-free and noise-free received signal, i.e.,
\begin{align}
{y}_{n,k}(t)=\sqrt{P\rho_{n}}b_ns_{n}(t-\tau_{n,k})e^{j2\pi f_{n,k}t}.
\end{align}
Define the signal-to-clutter-plus-noise ratio (SCNR) as $\delta_{\rm cn}=\frac{P}{T_{\rm eff}\sigma_{\rm cn}^2}$ with $T_{\rm eff}$ denoting the effective waveform width.

\textbf{Communication  model: } The received signal of the communication user is given by 
\begin{align}
\small
   {r}_{u}(t) &= \sqrt{P}\sum\nolimits_{n=1}^Ng_{n}{b}_{n}\sqrt{\rho_{n}}s_n(t)+{z}(t),\label{eq:received sig commun}  
\end{align}
where $g_{n}$ is the channel fading\footnote{The communication channels are assumed to be estimated a priori using pilot signals \cite{Behdad2024, Huang2023, Ahmed2019, Ahmed2024}.} between the transmitter $n$ and the user, and $z(t)$ denotes the interference-plus-noise with zero mean and variance $\sigma_{z}^2$. Then, the SINR at the user terminal can be given by
\begin{align}
\small
   \gamma &= \sum\nolimits_{n=1}^N|g_{n}{b}_{n}|^2\rho_{n}\delta\overset{(\text{i})}{=} \delta{\bm \rho}^{\rm T}{\bm g}\label{eq:SINR}  
\end{align}
where $\delta=\frac{P}{T_{\rm eff}\sigma_{z}^2}$ is the total loss-free SINR and ${\bm g}$ is the squared channel fading with its $n$th element being $|g_{n}|^2$. Note that $(\mathrm i)$ holds when using a normalized conjugate precoding with ${b}_{n}={g}^*_{n}/|g_{n}|$.

Because both sensing and communication channels are assumed quasi-static/deterministic within the coherence observation time, the deterministic RCS model and the estimated communication gains remain valid for each optimization cycle.

\vspace{-1em}
\section{The Proposed Sensing Metric and Problem Formulation}\label{sec: Metric and Problem}
In this section, we propose a sensing metric of the joint location and velocity estimation based on our derived approximate CRLB for CF MIMO radar under arbitrary PAs. 

To begin with, we introduce the following definitions. Define the squared effective bandwidth (SEBW) and squared effective pulse time width (SETW) as
\begin{align} \small
   \bar{f_n^2}=\frac{1}{4\pi^2}\int \bigg|\frac{d{s_{n}(t-\tau_{n,k})}}{d{\tau_{n,k}}}\bigg|^2\, dt=\int f^2|S_n(f)|^2\, df,\label{eq:square bandwidth}
\end{align}
and $ \bar{t_n^2}=\int t^2|s_n(t)|^2\, dt$, 
respectively. The cross-term between time and frequency is further defined as 
\begin{align} \small
\sigma_{tf}=\int{ts_{n}^*(t-\tau_{n,k})}\frac{d{s_{n}(t-\tau_{n,k})}}{d{\tau_{n,k}}}\, dt.  \label{eq:cross term}
\end{align} 
$|s_n(t)|^2$ and $|S_n(f)|^2$ can be interpreted as the probability density functions for consecutive random variables $t$ and $f$, respectively, 
since $\int |S_n(f)|^2\ df=\int |s_n(t)|^2\ dt=1$.  Based on this interpretation, we define the average time and average frequency of the transmission waveform as 
\begin{align} \small
\bar{f}_n=\int f|S_n(f)|^2\ df \quad 
 \text{and} \quad 
    \bar{t}_n=\int t|s_n(t)|^2\ dt,  \label{eq:average time}
\end{align}
respectively.


\vspace{-0.5em}
\subsection{Sensing Metric}
Unlike the sensing performance metrics employed in \cite{Behdad2024, Fan2025, Ahmed2019,Huang2023, Ahmed2024}, which only consider target detection or location estimation, we consider the joint location and velocity CRLB as the sensing metric, developed based on the \emph{discrete-time reception} of the reflected signal under the \emph{deterministic RCS assumption}. 
We derive the approximate CRLBs of the joint location and velocity estimation\footnote{For completeness, the detailed derivations are referred to~\cite{xia2025cell}.}, respectively, given by 
\begin{align} \small
   {\bm C}_{\rm L}& = ({\bm P}-{\bm V}{\bm Y}^{-1}{\bm V}^{\rm T})^{-1}, \label{eq:CRLB location}
\end{align}
and 
\begin{align} \small
   {\bm C}_{\rm V} &= ({\bm Y}-{\bm V}^{\rm T}{\bm P}^{-1}{\bm V})^{-1}.\label{eq:CRLB velocity}
\end{align}
where ${\bm P}$ and  ${\bm V}$ are presented at the top of the page, respectively, and 
\begin{figure*}
   \small
\begin{align}
   {\bm P}=\sum_{n=1}^N\sum_{k=1}^K\begin{bmatrix}\beta_{n,k}^2a_{n,k}+2\beta_{n,k}\eta_{n,k}b_{n,k}+\eta_{n,k}^2d_{n,k}&\beta_{n,k}\zeta_{n,k}a_{n,k} \!+ \!(\beta_{n,k}\kappa_{n,k} \!+ \!\zeta_{n,k}\eta_{n,k})b_{n,k}+\eta_{n,k}\kappa_{n,k}d_{n,k}\\\zeta_{n,k}\beta_{n,k}a_{n,k} \!+ \!(\zeta_{n,k}\eta_{n,k}+\beta_{n,k}\kappa_{n,k})b_{n,k}+\kappa_{n,k}\eta_{n,k}d_{n,k}&\zeta_{n,k}^2a_{n,k}+2\zeta_{n,k}\kappa_{n,k}b_{n,k}\!+ \! \kappa_{n,k}^2d_{n,k}
   \end{bmatrix}, \notag
\end{align}
\vspace{-3em}
\end{figure*}
\begin{figure*} 
\small
\begin{align} \small
{\bm V}= \sum_{n=1}^N\sum_{k=1}^K\begin{bmatrix}\beta_{n,k}\xi_{n,k}b_{n,k}+\eta_{n,k}\xi_{n,k}d_{n,k}&\beta_{n,k}\varrho_{n,k}b_{n,k}+\eta_{n,k}\varrho_{n,k}d_{n,k}\\\zeta_{n,k}\xi_{n,k}b_{n,k}+\kappa_{n,k}\xi_{n,k}d_{n,k}&\zeta_{n,k}\varrho_{n,k}b_{n,k}+\kappa_{n,k}\varrho_{n,k}d_{n,k}\end{bmatrix},\notag
\end{align}
 {\noindent}\rule[-10pt]{18cm}{0.05em}
\vspace{-1em}
\end{figure*}
\begin{align} \small
{\bm Y}= \sum_{n=1}^N\sum_{k=1}^K\begin{bmatrix}\xi_{n,k}^2d_{n,k}&\xi_{n,k}\varrho_{n,k}d_{n,k}\\\varrho_{n,k}\xi_{n,k}d_{n,k}&\varrho_{n,k}^2d_{n,k}\end{bmatrix}.\label{eq: mat Y}
\end{align}
The waveform-related parameters $a_{n,k}$, $b_{n,k}$ and $d_{n,k}$, and geometric-spreading-related factors $\beta_{n,k}$, $\zeta_{n,k}$, $\xi_{n,k}$, $\varrho_{n,k}$, $\eta_{n,k}$, and $\kappa_{n,k}$, are detailed in Appendix \ref{sec: CRLB details}.

As described in \eqref{eq:SINR}, the PA coefficients directly affect communication SINR. \emph{However, the PA optimization relevant to the CRLB is intractable due to the implicit and complex forms of the expressions in \eqref{eq:CRLB location} and \eqref{eq:CRLB velocity}}. To facilitate analysis and optimization, we reformulate the CRLB expressions into a more explicit and tractable form regarding the PA coefficients.

We denote the $(i,j)$th elements of matrices ${\bm P}$, ${\bm V}$ and ${\bm Y}$ by $p_{ij}$, $v_{ij}$ and $y_{ij}$, respectively. 
They can be rewritten as
\begin{align} \small
   \begin{cases}
   p_{ij}=\sum_{n=1}^N w_{n}^{p_{ij}}\rho_{n}={\bm \rho}^{\rm T}{\bm w}_{p_{ij}},\\
   v_{ij}=\sum_{n=1}^N w_{n}^{v_{ij}}\rho_{n}={\bm \rho}^{\rm T}{\bm w}_{v_{ij}},\\
   y_{ij}=\sum_{n=1}^N w_{n}^{y_{ij}}\rho_{n}={\bm \rho}^{\rm T}{\bm w}_{y_{ij}},
\end{cases}\label{eq:weighted sum}
\end{align} 
where $w_{n}^{p_{ij}}$, $w_{n}^{v_{ij}}$ and $w_{n}^{y_{ij}}$ are the $n$th element of the weight vector ${\bm w}_{p_{ij}}$, ${\bm w}_{v_{ij}}$ and ${\bm w}_{y_{ij}}$, respectively. These weights can be readily obtained by extracting $\rho_n$ from the element-wise definitions of ${\bm P}$, ${\bm V}$ and ${\bm Y}$. For example, we have
\begin{align} \small
w_{n}^{p_{11}}=\sum\nolimits_{k=1}^K(\beta_{n,k}^2\bar{a}_{n,k}+2\beta_{n,k}\eta_{n,k}\bar{b}_{n,k}+\eta_{n,k}^2\bar{d}_{n,k}), 
\end{align} 
where the details of $\bar{a}_{n,k}$, $\bar{b}_{n,k}$ and $\bar{d}_{n,k}$ are referred to \eqref {eq:simplified SO-PD Delay_Doppler}.

By substituting \eqref{eq:weighted sum} into \eqref{eq:CRLB location}, the CRLB matrix for location estimation can be rewritten as
\begin{align} \small
   {\bm C}_{\rm L}=\frac{1}{|{\bm U}|}\begin{bmatrix}u_{22}&-u_{12}\\-u_{21}&u_{11}
   \end{bmatrix}, \label{eq:CRLB location U}
\end{align}
where $|{\bm U}|=u_{11}u_{22}-u_{12}u_{21}$. 
The elements $u_{ij}$ are given by
\begin{equation}
\small
   \begin{aligned} 
      u_{11} = 
      \frac{1}{|{\bm Y}|}\big[{\bm \rho}^{\rm T}({\bm w}_{p_{11}}{\bm \rho}^{\rm T}{\bm W}_y\!-\!{\bm w}_{v_{11}}{\bm \rho}^{\rm T}{\bm W}_{12}^y\!-\!{\bm w}_{v_{12}}{\bm \rho}^{\rm T}\breve{{\bm W}}_{11}^y){\bm \rho}\big],\\
      u_{12}=\frac{1}{|{\bm Y}|}\big[{\bm \rho}^{\rm T}({\bm w}_{p_{12}}{\bm \rho}^{\rm T}{\bm W}_y\!-\!{\bm w}_{v_{21}}{\bm \rho}^{\rm T}{\bm W}_{12}^y\!-\!{\bm w}_{v_{22}}{\bm \rho}^{\rm T}\breve{{\bm W}}_{11}^y){\bm \rho}\big],\\
      u_{22}=\frac{1}{|{\bm Y}|}\big[{\bm \rho}^{\rm T}({\bm w}_{p_{22}}{\bm \rho}^{\rm T}{\bm W}_y \! - \!{\bm w}_{v_{21}}{\bm \rho}^{\rm T}{\bm W}_{22}^y \!- \!{\bm w}_{v_{22}}{\bm \rho}^{\rm T}\breve{{\bm W}}_{21}^y){\bm \rho}\big],
   \end{aligned} \label{eq:element U simplified}
\end{equation}
where $|{\bm Y}|={\bm \rho}^{\rm T}{\bm W}_y{\bm \rho}$, and 
\begin{align} \small
   \begin{cases}
      {\bm W}_y={\bm w}_{y_{11}}{\bm w}_{y_{22}}^{\rm T}-{\bm w}_{y_{12}}{\bm w}_{y_{21}}^{\rm T},\\
      {\bm W}^y_{12}={\bm w}_{v_{11}}{\bm w}_{y_{22}}^{\rm T}-{\bm w}_{v_{12}}{\bm w}_{y_{21}}^{\rm T},\\
      {\bm W}^y_{22}={\bm w}_{v_{21}}{\bm w}_{y_{22}}^{\rm T}-{\bm w}_{v_{22}}{\bm w}_{y_{21}}^{\rm T},\\
      \breve{\bm W}^y_{11}={\bm w}_{v_{12}}{\bm w}_{y_{11}}^{\rm T}-{\bm w}_{v_{11}}{\bm w}_{y_{12}}^{\rm T},\\
      \breve{\bm W}^y_{21}={\bm w}_{v_{22}}{\bm w}_{y_{11}}^{\rm T}-{\bm w}_{v_{21}}{\bm w}_{y_{12}}^{\rm T}.
   \end{cases} \label{eq:Para Mat W_y}
\end{align}
Similarly, the velocity CRLB matrix can be rewritten as 
\begin{align} \small
   {\bm C}_{\rm V}=\frac{1}{|{\bm H}|}\begin{bmatrix}h_{22}&-h_{12}\\-h_{21}&h_{11}
   \end{bmatrix},\label{eq:CRLB velocity H}
\end{align}
where $|{\bm H}|=h_{11}h_{22}-h_{12}h_{21}$. The relevant elements $h_{ij}$ are given respectively by 
\begin{equation}
\small
   \begin{aligned} \small
      h_{11}=\frac{1}{|{\bm P}|}\big[{\bm \rho}^{\rm T}({\bm w}_{y_{11}}{\bm \rho}^{\rm T}{\bm W}_p\!-\!{\bm w}_{v_{11}}{\bm \rho}^{\rm T}{\bm W}_{12}^p\!-\!{\bm w}_{v_{21}}{\bm \rho}^{\rm T}\breve{{\bm W}}_{11}^p){\bm \rho}\big],\\
      h_{12}=\frac{1}{|{\bm P}|}\big[{\bm \rho}^{\rm T}({\bm w}_{y_{12}}{\bm \rho}^{\rm T}{\bm W}_p\!-\!{\bm w}_{v_{12}}{\bm \rho}^{\rm T}{\bm W}_{12}^p\!-\!{\bm w}_{v_{22}}{\bm \rho}^{\rm T}\breve{{\bm W}}_{11}^p){\bm \rho}\big],\\
      h_{22}=\frac{1}{|{\bm P}|}\big[{\bm \rho}^{\rm T}({\bm w}_{y_{22}}{\bm \rho}^{\rm T}{\bm W}_p\!-\!{\bm w}_{v_{12}}{\bm \rho}^{\rm T}{\bm W}_{22}^p\!-\!{\bm w}_{v_{22}}{\bm \rho}^{\rm T}\breve{{\bm W}}_{21}^p){\bm \rho}\big],
   \end{aligned} \label{eq:element H simplified}
\end{equation}
where $|{\bm P}|={\bm \rho}^{\rm T}{\bm W}_p{\bm \rho}$, and
\begin{align} \small
   \begin{cases}
      {\bm W}_p={\bm w}_{p_{11}}{\bm w}_{p_{22}}^{\rm T}-{\bm w}_{p_{12}}{\bm w}_{p_{21}}^{\rm T},\\
      {\bm W}^p_{12}={\bm w}_{v_{11}}{\bm w}_{p_{22}}^{\rm T}-{\bm w}_{v_{21}}{\bm w}_{p_{21}}^{\rm T},\\
      {\bm W}^p_{22}={\bm w}_{v_{12}}{\bm w}_{p_{22}}^{\rm T}-{\bm w}_{v_{22}}{\bm w}_{p_{21}}^{\rm T},\\
      \breve{\bm W}^p_{11}={\bm w}_{v_{21}}{\bm w}_{p_{11}}^{\rm T}-{\bm w}_{v_{11}}{\bm w}_{p_{12}}^{\rm T},\\
      \breve{\bm W}^p_{21}={\bm w}_{v_{22}}{\bm w}_{p_{11}}^{\rm T}-{\bm w}_{v_{12}}{\bm w}_{p_{12}}^{\rm T}.
   \end{cases} \label{eq:Para Mat W_p}
\end{align}
The newly derived CRLB expressions in \eqref{eq:CRLB location U} and \eqref{eq:CRLB velocity H} represent explicit functions of the PA coefficient vector ${\bm \rho}$, thereby serving as effective and tractable sensing performance metrics.

\vspace{-1em}
\subsection{Problem Formulation}\label{sec:Problem Formulation}
This section formulates the PA problem by maximizing the overall performance of CF MIMO-ISAC systems. Depending on specific application requirements, the optimization objective can be either sensing or communication performance, while the requirement of the other functionality is incorporated as a constraint in the optimization model.

This paper considers the problem of maximizing the communication SINR in \eqref{eq:SINR}, subject to constraints on the performance of joint location and velocity estimation,  i.e.,
\begin{subequations}
\begin{align} \small
   \arg\min\limits_{{\bm \rho}}&\ -{\bm \rho}^{\rm T}{\bm g},\label{eq:problem}\\
   {\rm s.t.}\ & {\bm 1}_N^{\rm T}{\bm \rho}=1, \label{eq:total power constraint} \\
      \ &{\bm \rho}_{\min}\leq{\bm \rho}\leq {\bm \rho}_{\max},\label{eq:individual power constraint}\\
      \ &{\rm tr}\{{\bm C}_{\rm L}\}\leq \delta_l^2,\label{eq:location constraint}\\
      \ &{\rm tr}\{{\bm C}_{\rm V}\}\leq \delta_v^2,\label{eq:velocity constraint}
\end{align}  
\end{subequations}
where $\delta_l^2$ and $\delta_v^2$ denote the corresponding location and velocity CRLB thresholds \footnote{The thresholds can be specified by referring to the CRLB values under uniform PA.}. 

Note that ${\bm \rho}_{\min} = [\rho_{\min,1}, \rho_{\min,2}, \ldots, \rho_{\min,N}]^{\mathrm{T}}$ denote the minimum PA coefficient vector required to ensure a sufficiently high SNR for achieving tight CRLBs~\cite{Ahmed2019, Ahmed2024}, and ${\bm \rho}_{\max} = [\rho_{\max,1}, \rho_{\max,2}, \ldots, \rho_{\max,N}]^{\mathrm{T}}$ denote the maximum PA coefficient vector determined by the individual transmitters’ power budgets~\cite{Godrich2011, Ahmed2024}.
Specifically, \eqref{eq:total power constraint} and \eqref{eq:individual power constraint} represent the total power and per-transmitter power constraints, respectively, while \eqref{eq:location constraint} and \eqref{eq:velocity constraint} correspond to the location and velocity CRLB constraints, respectively.

\vspace{-1em}
\section{Power Allocation Algorithm for CF MIMO-ISAC}\label{sec:algorithm derivation}

We now investigate the PA strategy by solving the optimization model formulated in \eqref{eq:problem}–\eqref{eq:velocity constraint}. The strong nonlinearity and nonconvexity of the CRLB-based constraints in \eqref{eq:location constraint} and \eqref{eq:velocity constraint} make the problem analytically intractable. Although numerous numerical optimization methods developed in the literature in past several decades, these methods cannot be directly applied to the nonlinear and nonconvex constrained PA optimization problem considered in this work.
To address this challenge, we develop a penalty function and projection operator enhanced conjugate gradient (CG)–based iterative algorithm, to yield an efficient numerical solution.
\vspace{-1em}
\subsection{Unconstrained Optimization Enabled by Penalty Function and Projection Operator}
To facilitate optimization, we start from converting the original constrained problem into an unconstrained one. Firstly, we consider incorporating the nonlinear inequality constraints in \eqref{eq:individual power constraint}-\eqref{eq:velocity constraint} onto the target function \eqref{eq:problem}, by introducing a penalty (function) for violating these constraints.
It can be defined as \cite{John2006Chapter9} 
\begin{equation} 
       \alpha({\bm \rho}) = \tilde{f}_l({\bm \rho})+\tilde{f}_v({\bm \rho})+\sum\nolimits_{n=1}^N\left(f_{\min,n}({\bm \rho})+f_{\max,n}({\bm \rho})\right),\notag
\end{equation}
where the involved penalty sub-functions are 
\begin{align} \small
   \begin{cases}
   \tilde{f}_l({\bm \rho}) = (\max\{0,{f}_l({\bm \rho})\})^q,\\
   \tilde{f}_v({\bm \rho}) = (\max\{0,{f}_v({\bm \rho})\})^q,\\
   f_{\min,n}({\bm \rho}) = (\max\{0,\rho_{\min,n}-\rho_n\})^q,\\
   f_{\max,n}({\bm \rho}) = (\max\{0,\rho_n-\rho_{\max,n}\})^q,
   \end{cases}
\end{align}
with ${f}_l({\bm \rho})\triangleq {\rm tr}\{{\bm C}_{\rm L}\}-\delta_l^2$ and ${f}_v({\bm \rho})\triangleq {\rm tr}\{{\bm C}_{\rm V}\}-\delta_v^2$. $q$ denotes the power to the penalty function, which is set to $q = 2$ in this work to ensure continuous differentiability.
Then, the penalty-driven objective function is given by
\begin{align} \small
   L({\bm \rho})=-{\bm \rho}^{\rm T}{\bm g}+\mu\alpha({\bm \rho}),\label{eq:target with penalty}
\end{align}
where $\mu$ denotes the penalty factor. 
Additionally, to reduce both computational complexity and the nonlinearity of the optimization problem, we adopt a projection operator to enforce the equality constraint \eqref{eq:total power constraint} instead of incorporating another penalty sub-function, i.e.,
\begin{align} \small
   {\bm \rho}({\bm \varpi})={\bm \Theta}{\bm \varpi}+{\bm \rho}_0,\label{eq:projection operator}
\end{align}
where ${\bm \varpi}\in\mathbb{R}^{N \times 1}$ is an arbitrary vector and ${\bm \Theta}\triangleq {\bm I}_{N\times N}-\frac{{\bm 1}_N{\bm 1}_N^{\rm T}}{N}$ is a orthogonal projection matrix onto the null space of ${\bm 1}_N^{\rm T}$, i.e., ${\bm 1}_N^{\rm T}{\bm \Theta}={\bm 0}_N^{\rm T}$. Thus, the equality constraint ${\bm 1}_N^{\rm T}{\bm \rho}({\bm \varpi})=1$ in \eqref{eq:total power constraint} always holds provided that ${\bm 1}_N^{\rm T}{\bm \rho}_0=1$. Consequently, ${\bm \rho}({\bm \varpi})$ is the sum of a constant PA vector ${\bm \rho}_0$ and an orthogonal projection of ${\bm \varpi}$ onto the null space of ${\bm 1}_N^{\rm T}$.  

Substituting \eqref{eq:projection operator} into \eqref{eq:target with penalty} can convert the original constrained optimization to an unconstrained problem as follows:
\begin{align} \small
   \arg\min\limits_{{\bm \rho}}&\ \mathcal{L} ({\bm \varpi})=-{\bm \rho}^{\rm T}({\bm \varpi}){\bm g}+\mu\alpha({\bm \rho}({\bm \varpi})). \label{eq:unconstrained optimization}
\end{align} 
Building upon the formulated problem, the next section presents unconstrained optimization methods for efficiently finding the PA solution.

\vspace{-1em}
\subsection{Modified Conjugate Gradient Algorithm}
To numerically solve \eqref{eq:unconstrained optimization}, we employ an iterative search algorithm. The $(i+1)$th update is given by
\begin{align} \small
   {\bm \varpi}_{i+1} = {\bm \varpi}_{i} + \upsilon_i{\bm d}_{i},\label{eq:intermediate parameter update}
\end{align}
where ${\bm d}_{i}$ is the search direction of ${\bm \varpi}_i$ and $\upsilon_i$ is the step size. For second-order convergence, we consider using a CG algorithm, with the search direction updated by
\begin{align}
   {\bm d}_{i+1} &= -\nabla\mathcal{L} ({\bm \varpi}_{i+1})+\varsigma_i{\bm d}_i, \label{eq:CG gradient update}
\end{align}
where $\varsigma_{i}$ denotes the deflection factor of a CG algorithm.
$\nabla\mathcal{L} ({\bm \varpi}_{i})$ is the derivative $\mathcal{L} ({\bm \varpi})$ regarding ${\bm \varpi}$, given by
\begin{align} \small
   \nabla\mathcal{L} ({\bm \varpi})=\ &{\bm \Theta}\nabla L({\bm \rho})={\bm \Theta}\big[{\bm g}+\mu\nabla\alpha({\bm \rho})\big],\label{eq:gradient}
\end{align}
where 
\begin{align} \small
\nabla\alpha({\bm \rho}) = \ &\mu\big(\nabla\tilde{f}_l({\bm \rho})+\nabla\tilde{f}_v({\bm \rho})\notag\\
   &+\sum\nolimits_{n=1}^N(\nabla f_{\min,n}({\bm \rho})+\nabla f_{\max,n}({\bm \rho}))\big),\label{eq:gradient component}
\end{align}
and
\begin{align} \small
   \begin{cases}
      \nabla\tilde{f}_l({\bm \rho})=2\max\{f_l({\bm \rho}),0\}\nabla f_l({\bm \rho})\\
      \nabla\tilde{f}_v({\bm \rho})=2\max\{f_v({\bm \rho}),0\}\nabla f_v({\bm \rho})\\
      \nabla f_{\min,n}({\bm \rho})=2\min\{\rho_n-\rho_{\min,n},0\}{\bm e}_n\\
      \nabla f_{\max,n}({\bm \rho})=2\max\{\rho_n-\rho_{\max,n},0\}{\bm e}_n. 
   \end{cases}
\end{align} 
Note that ${\bm e}_n\in \mathbb{R}^{N\times 1}$ is an index vector with its element at index $n$ being $1$ and the others being zeros. 
Different from classic CG algorithm, \eqref{eq:gradient} is a projected gradient direction regarding ${\bm \rho}$.  

In the following, we will directly solve PA vector ${\bm \rho}$ instead of updating ${\bm \varpi}$. By using the idempotent property of the projection matrix, i.e., ${\bm \Theta}^2={\bm \Theta}$, we firstly have ${\bm \Theta}\nabla\mathcal{L} ({\bm \varpi})=\nabla\mathcal{L} ({\bm \varpi})$ for the gradient $\nabla\mathcal{L} ({\bm \varpi})$ in \eqref{eq:gradient}. Then, by setting the initial direction ${\bm d}_{0} = -{\bm \Theta}\nabla L({\bm \rho}_0)$ in \eqref{eq:CG gradient update}, we further have ${\bm \Theta}{\bm d}_{i}={\bm d}_{i}$, $\forall{i>0}$.
Finally, taking the projection of \eqref{eq:intermediate parameter update} leads to 
\begin{align} \small
   {\bm \rho}_{i+1} = {\bm \rho}_{i} + \upsilon_i{\bm d}_{i},\label{eq:parameter update}
\end{align}
where the search direction is updated by
\begin{align} \small
   {\bm d}_{i+1} &= -{\bm \Theta}\nabla\mathcal{L} ({\bm \rho}_{i+1})+\varsigma_i{\bm d}_i.\label{eq:search direction update}
\end{align} 
We refer to \eqref{eq:parameter update} and \eqref{eq:search direction update} as the penalty-function and projection-based modified CG method (PP-MCG). Different from \eqref{eq:intermediate parameter update}, \eqref{eq:parameter update} yields the numerically iterative solution of the penalty-driven objective function \eqref{eq:target with penalty}.
Additionally, with the search direction in \eqref{eq:search direction update}, the equality constraint \eqref{eq:total power constraint}, that is, ${\bm 1}_N^{\rm T}{\bm \rho}_{i}=1$, always holds at the $i$th iteration for $\forall{i>0}$ if and only if the initial PA vector ${\bm \rho}_0$ satisfies ${\bm 1}_N^{\rm T}{\bm \rho}_0=1$. This is because ${\bm 1}_N^{\rm T}{\bm d}_{i}=0$ for $\forall{i>0}$ provided that ${\bm 1}_N^{\rm T}{\bm d}_0=0$, due to ${\bm 1}_N^{\rm T}{\bm \Theta}={\bm 0}_N^{\rm T}$. 

\vspace{-1em}
\subsection{Parameter Optimization of PP-MCG Algorithm} 
A necessary condition for the convergence of the PP-MCG algorithm is that the objective function is non-increasing at each iteration, i.e.,
\begin{align} \small
L({\bm \rho}_{i} + \upsilon_i{\bm d}_{i}) - L({\bm \rho}_{i}) \leq 0. \label{eq:descent function}
\end{align}
To this end, ${\bm d}_{i}$ should be a \emph{descent direction}, and the \emph{step size} $\upsilon_i$ should be optimized at each iteration. To ensure the descent direction, the \emph{deflection factor} $\varsigma_i$ is crucial.

\subsubsection{Descent direction}
To ensure that ${\bm d}_{i+1}$ is a descent direction of $L({\bm \rho}_{i+1})$, it is required that
 \begin{align} \small
    {\bm d}_{i+1}^{\rm T}\nabla L({\bm \rho}_{i+1})\leq 0,  \label{eq:search direction constraint}
 \end{align}
where the equality holds if and only if $\nabla L({\bm \rho}_{i+1})={\bm 0}$. The initial direction ${\bm d}_{0} = -{\bm \Theta}\nabla L({\bm \rho}_0)$ is descent with ${\bm d}_{0}^{\rm T}\nabla L({\bm \rho}_0)\leq 0$ since the projection operator ${\bm \Theta}$ is a symmetric positive definite matrix. 
By substituting \eqref{eq:search direction update} into \eqref{eq:search direction constraint}, the feasible range of the deflection factor can be obtained, i.e.,
\begin{align} \small
   \begin{cases}
      \varsigma_i\leq \varsigma_i^{\rm b}, \ \varsigma_i^{\rm b}>0 \\
      \varsigma_i\geq \varsigma_i^{\rm b}, \ \varsigma_i^{\rm b}<0
   \end{cases}\label{eq:deflection constraint}
\end{align}
 where $\varsigma_i^{\rm b}$ denotes the bound in selecting $\varsigma_i$, given by
\begin{align} \small
\varsigma_i^{\rm b}=\dfrac{\nabla L({\bm \rho}_{i+1})^{\rm T}{\bm \Theta}\nabla L({\bm \rho}_{i+1})}{{\bm d}_i^{\rm T}\nabla L({\bm \rho}_{i+1})}.
\end{align}
Note that $\nabla L({\bm \rho}_{i+1})^{\rm T}{\bm \Theta}\nabla L({\bm \rho}_{i+1})>0$ for any nonzero gradient $\nabla L({\bm \rho}_{i+1})$, owing to the positive definiteness of ${\bm \Theta}$. 

When employing inexact line searches, the deflection selection for $\varsigma_i$ proposed by Hestenes and Stiefel \cite{Hestenes1952}, referred to as HS method, is preferable.
Herein, we modify the HS deflection for the projection-driven iterative method, i.e.,
\begin{align} \small
   \varsigma_i^{\rm HS} = {\nabla L({\bm \rho}_{i+1})^{\rm T}{\bm q}_i}/{{\bm d}_i^{\rm T}{\bm q}_i}, \label{eq:HS factor}
\end{align}
which is obtained from ${\bm d}_{i+1}^{\rm T}{\bm q}_i=0$ with ${\bm q}_i\triangleq {\bm \Theta}[\nabla L({\bm \rho}_{i+1})-\nabla L({\bm \rho}_{i})]$. 

In fact, when using the MCG algorithm, it is desirable to maintain orthogonality between the conjugate gradients of adjacent iterations, i.e., 
\begin{align} \small
&|{\bm d}_i^{\rm T}\nabla L({\bm \rho}_{i+1})|\ll |{\bm d}_i^{\rm T}\nabla L({\bm \rho}_{i})|,\label{eq:conjugacy 1}\\
&|\nabla L({\bm \rho}_{i+1})^{\rm T}{\bm \Theta}\nabla L({\bm \rho}_{i})|\ll |\nabla L({\bm \rho}_{i+1})^{\rm T}{\bm \Theta}\nabla L({\bm \rho}_{i+1})|. \label{eq:conjugacy 2}
\end{align} 
Note that \eqref{eq:conjugacy 1} and \eqref{eq:conjugacy 2} ensure the \emph{positivity} of ${\bm d}_i^{\rm T}{\bm q}_i$ and $\nabla L({\bm \rho}_{i+1})^{\rm T}{\bm q}_i$, respectively. This allows for a positive conjugate gradient deflection factor $\varsigma_i^{\rm HS}$.
Furthermore, \eqref{eq:conjugacy 1} and \eqref{eq:conjugacy 2} also imply that $|{\bm d}_i^{\rm T}\nabla L({\bm \rho}_{i+1})|\ll |{\bm d}_i^{\rm T}{\bm q}_i|$ and $|\nabla L({\bm \rho}_{i+1})^{\rm T}{\bm q}_i|\approx |\nabla L({\bm \rho}_{i+1})^{\rm T}{\bm \Theta}\nabla L({\bm \rho}_{i+1})|$, respectively.
As a result, $\varsigma_i^{\rm HS}$ is expected to be smaller than $|\varsigma_i^{\rm b}|$, thereby satisfying the deflection constraint defined in \eqref{eq:deflection constraint}. 

Rather than strictly relying on the potentially tight and unquantifiable conjugacy conditions in \eqref{eq:conjugacy 1} and \eqref{eq:conjugacy 2}, we instead directly enforce the \emph{positivity} of both the numerator and denominator in \eqref{eq:HS factor}, along with the inequality $\varsigma_i^{\rm HS} < |\varsigma_i^{\rm b}|$. When these conditions hold, we set $\varsigma_i = \varsigma_i^{\rm HS}$. Otherwise, we reset the MCG iteration by letting $\varsigma_i = 0$. \emph{To sum up, the deflection factor is updated by}
\begin{equation}
\small
   {\varsigma_i=}
   \begin{cases}
      \varsigma_i^{\rm HS},\ \min\{\nabla L({\bm \rho}_{i+1})^{\rm T}{\bm q}_i,{\bm d}_i^{\rm T}{\bm q}_i,|\varsigma_i^{\rm b}|-\varsigma_i^{\rm HS}\}>0,\\
      0,\ {\rm else}.
   \end{cases}\label{eq:deflection choice}
\end{equation}
Note that, after every consecutive $Q$ MCG steps with $\varsigma_i=\varsigma_i^{\rm HS}$, we reset $\varsigma_i=0$ to initiate a new cycle of MCG iterations. Such a reboot strategy can help mitigate the accumulation of numerical errors, restoring the orthogonality and conjugacy of direction vectors, and avoiding local minima in nonlinear and non-convex optimization.

\subsubsection{Step size optimization}
We have explored the selection of deflection factor above to ensure a descent direction in \eqref{eq:search direction constraint}. The second step is to
optimize the step size selection. 

Let us define $f_i(\upsilon_i)\triangleq L({\bm \rho}_{i} + \upsilon_i{\bm d}_{i})$ with $f_i(0)=L({\bm \rho}_{i})$. The second-order Taylor approximation of $f_i(\upsilon_i)$ at $\upsilon_i=0$ is given by
\begin{align} \small
   f_i(\upsilon_i) &\thickapprox  f_i(0)+\upsilon_i\nabla f_i(0)+\upsilon_i^2\nabla^2 f_i(0)/2\notag\\
   &= f_i(0)+\upsilon_i{\bm d}_{i}^{\rm T}\nabla L({\bm \rho}_{i})+\upsilon_i^2/2{\bm d}_{i}^{\rm T}\nabla^2L({\bm \rho}_{i}){\bm d}_{i}, \label{eq:SO approx}
\end{align} 
where $\nabla^2L({\bm \rho}_{i})$ is the Hessian matrix of $L({\bm \rho})$ at ${\bm \rho}_{i}$. Using the optimized deflection factor above, we have ${\bm d}_{i}^{\rm T}\nabla L({\bm \rho}_{i})\leq 0$. 
If $\|\nabla^2L({\bm \rho}_{i})\|$ is locally bounded, there will exist a non-negative step size bound $\upsilon_{i,\rm b}$ for a given constant $\epsilon_{L}\in[0, 1)$ such that
\begin{align} \small
f_i(\upsilon_i)-f_i(0)\leq\epsilon_{L} \upsilon_i{\bm d}_{i}^{\rm T}\nabla L({\bm \rho}_{i}),\quad   \text {for} \quad    \upsilon_i\leq\upsilon_{i,\rm b}.    \label{eq:Pre Armijo rule}
\end{align} 
When ${\bm d}_{i}^{\rm T}\nabla^2L({\bm \rho}_{i}){\bm d}_{i}\leq 0$, \eqref{eq:Pre Armijo rule} always holds. Otherwise, we have $\upsilon_{i,\rm b}=\dfrac{2(\epsilon_{L}-1){\bm d}_{i}^{\rm T}\nabla L({\bm \rho}_{i})}{{\bm d}_{i}^{\rm T}\nabla^2L({\bm \rho}_{i}){\bm d}_{i}}$ when ${\bm d}_{i}^{\rm T}\nabla^2L({\bm \rho}_{i}){\bm d}_{i}>0$. One can choose a step size at each iteration as $\upsilon_i=\delta_i\upsilon_{i,\rm b}$ for the latter case, with a small constant $\delta_i\in(0,1)$.

However, in this work, due to the high computational complexity of the Hessian matrix $\nabla^{2}L({\bm \rho}_{i})$ when computing $\upsilon_{i,\mathrm{b}}$, we adopt an inexact line search (ILS) strategy for step-size selection to satisfy \eqref{eq:Pre Armijo rule}, i.e.,
\begin{align} \small
   L({\bm \rho}_{i} + \upsilon_i{\bm d}_{i})-L({\bm \rho}_{i})\leq\epsilon_{L} \upsilon_i{\bm d}_{i}^{\rm T}\nabla L({\bm \rho}_{i}).\label{eq:Armijo rule}
\end{align}
Note that \eqref{eq:Armijo rule} can ensure a descent function in \eqref{eq:descent function} since the term on the right-hand side is always non-positive. 
The ILS controlling factor $\epsilon_{L}$ is used to regulate the decrement of the function $L({\bm \rho})$ by moving ${\bm \rho}_{i}$  towards the direction ${\bm d}_{i}$ weighted by the step size $\upsilon_i$. When \eqref{eq:Armijo rule} is unmet for a given $\upsilon_i$, the step size is scaled by 
\begin{align} \small
   \upsilon_i = \sigma_{\upsilon}\upsilon_i,
   \label{eq:step scaling}
\end{align} 
where $\sigma_{\upsilon}\in(0,1)$ is a scaling factor. An initial step size, denoted as $\upsilon_i=\upsilon_{i,0}$, can be chosen as described later.  

Additionally, negative PA values may occur in intermediate iterations. Such infeasible values can lead to misleading CRLB evaluations, e.g., producing negative rather than infinite CRLBs, and may thus falsely appear to satisfy the CRLB constraints. To prevent unnecessary iterations within the infeasible region and to ensure physically meaningful results, it is essential to enforce non-negative PAs at each iteration step. To this end, the step size $\upsilon_i$ needs to satisfy 
\begin{align} \small
   0<\upsilon_i\leq \upsilon_i^{\rm b},
\end{align}
where $\upsilon_i^{\rm b}$ denotes an upper bound on the step size, given by
\begin{align}
   \upsilon_i^{\rm b}=\min_{d_{n,i}<0}\{\rho_{n,i}/|{d}_{n,i}|\},\label{eq:step size range}
\end{align} 
with $\rho_{n,i}$ denoting the $n$th element of ${\bm \rho}_{i}$ and similar for ${d}_{n,i}$. Note that there must exist at least one component ${d}_{n,i}<0$ in each iteration $i$, unless ${\bm d}_{i}={\bm 0}$, due to the constraint ${\bm 1}_N^{\rm T}{\bm d}_{i}=0$. Thus, the initial step size for ILS can be selected as
\begin{align}  \small
\upsilon_{i,0}=\epsilon_{\upsilon}\upsilon_i^{\rm b}\label{eq:step size initial}
\end{align} 
with $\epsilon_{\upsilon}\in(0,1]$.  \emph{The ILS method}, characterized by \eqref{eq:Armijo rule}–\eqref{eq:step size initial}, can be regarded as an adaptation of Armijo’s rule~\cite{Armijo1966}, customized to handle the non-negative PA optimization under projection constraint in our proposed framework.
\subsubsection{The algorithm implementation}
A well-defined termination condition is essential for the proposed iterative algorithm. A natural choice, inspired by the ILS condition in \eqref{eq:Armijo rule}, is to monitor the absolute change in the objective function between consecutive iterations, i.e., $|L({\bm \rho}_{i+1})-L({\bm \rho}_{i})|\leq\epsilon$, where $\epsilon$ is a predefined small threshold. To avoid numerical instability arising from the magnitude of the objective function itself, we apply a normalized termination condition, i.e., $|L({\bm \rho}_{i+1})-L({\bm \rho}_{i})|/|L({\bm \rho}_{i+1})|\leq\epsilon_{\rm th}$. This is employed in both the MCG-based and the following MSD-based algorithms.

Finally, the selection of penalty factor is also important in guaranteeing a good learning performance. Generally, a large penalty factor may lead to numerical issues, such as an ill-conditioned Hessian matrix, which can degrade the convergence behavior of the MCG algorithm. To mitigate this, we adopt a sequential penalty strategy that gradually increases the penalty factor across successive iteration loops.
Specifically, for a given penalty factor $\mu_k$, the MCG algorithm is executed until its termination condition is satisfied. Then, the penalty factor is updated as $\mu_{k+1} = \varphi \mu_k$ where $\varphi > 1$ is a scaling constant, for the next loop of penalty-based iteration. This iterative refinement continues until the convergence criterion $\mu_k \alpha(\bm{\rho}) \leq \epsilon_{\mu}$ is satisfied, where  $\epsilon_{\mu}$  is a small, predefined threshold \cite{John2006Chapter9}.
This sequential penalty strategy ensures both numerical stability and progressive enforcement of the constraint conditions. The detailed procedure of the PP-MCG-ILS algorithm is summarized in \textbf{Algorithm}~\ref{alg:PP-MCG}.
\begin{algorithm}[!t]
   \small
	\renewcommand{\algorithmicrequire}{\textbf{Input:}}
	\renewcommand{\algorithmicensure}{\textbf{Output:}}
	\caption{The PP-MCG-ILS algorithm}
	\label{alg:PP-MCG}
   \textbf{Input:} The positive scalar termination factor $\epsilon_{\mu}$ and $\epsilon_{\rm th}$, small positive constant $\sigma_{\upsilon}$, $\epsilon_L$ and $\epsilon_{\upsilon}$, initial penalty factor $\mu_1$, scaling factor $\varphi>1$; initial PA parameter ${\bm \rho}_0$ satisfying ${\bm 1}^{\rm T}{\bm \rho}_0=1$, projection operator ${\bm \Theta}$, ${\bm y}_1={\bm \rho}_1$, and ${\bm d}_1= -{\bm \Theta}\nabla L({\bm \rho}_1)$; a counter $k=0$ and $i=j=1$.\\
   \textbf{Output:} The PA vector ${\bm y}_j$.
    \begin{algorithmic}[1]
          \STATE Step size bound: Calculate $\upsilon_i^{\rm b}$ according to \eqref{eq:step size range}; 
          \STATE Step size search: set $\upsilon_i=\epsilon_{\upsilon}\upsilon_i^{\rm b}$, while $L({\bm \rho}_{i} + \upsilon_i{\bm d}_{i})-L({\bm \rho}_{i})>\epsilon_L \upsilon_i{\bm d}_{i}^{\rm T}\nabla L({\bm \rho}_{i})$, let $\upsilon_i = \sigma_{\upsilon}\upsilon_i$. 
          \STATE PA update: ${\bm \rho}_{i+1}={\bm \rho}_{i}+\upsilon_i{\bm d}_{i}$.
          \STATE Convergence condition: If $|L({\bm \rho}_{i+1})-L({\bm \rho}_{i})|/|L({\bm \rho}_{i+1})|\leq\epsilon_{\rm th}$, ${\bm y}_j={\bm \rho}_{i+1}$, and go to \textbf{step 7}; Otherwise, go to next step.
          \STATE Deflection choice: According to \eqref{eq:deflection choice}, if $\varsigma_i=\varsigma_i^{\rm HS}$, $k=k+1$, or $k=0$ otherwise; when $k=Q$, let $\varsigma_i=0$ and $k=0$; 
          \STATE Direction update: ${\bm d}_{i+1}= -{\bm \Theta}\nabla L({\bm \rho}_{i+1})+\varsigma_i{\bm d}_i$; let $i=i+1$, and go back to \textbf{step 1}.
          \STATE Penalty update: If $\mu_j\alpha({\bm y}_j)<\epsilon_{\mu}$, stop; Otherwise, $\mu_{j+1}=\varphi\mu_j$, and ${\bm d}_{i+1} = -{\bm \Theta}\nabla L({\bm \rho}_{i+1})$; let $j=j+1$ and $i=i+1$, and go to \textbf{step 1}.
    \end{algorithmic}
\end{algorithm}

   \vspace{-1em}
\subsection{Modified Steepest Descent Algorithm}
In fact, alternative unconstrained optimization methods can be applied to solve the proposed projection-based penalty problem in \eqref{eq:unconstrained optimization}. In this paper, we also develop an MSD algorithm, adapted from the proposed MCG framework. The main difference between the MSD and MCG algorithms lies in how the search direction is determined. In the MSD approach, the search direction is set to the projected negative gradient of the penalty-based objective function, i.e.,
\begin{align} \small
   {\bm d}_i &= -{\bm \Theta}\nabla L({\bm \rho}_i),\label{eq:search direction MSD}
\end{align}
which can be obtained from MCG by setting the deflection factor to zero.
We herein omit the detailed implementation steps of PP-MSD-ILS since they can be obtained from the PP-MCG-ILS in 	\textbf{Algorithm} \ref{alg:PP-MCG} by simply removing \textbf{step 5} and performing the direction update in \textbf{step 6} using \eqref{eq:search direction MSD}.
   \vspace{-1em}
\subsection{Benchmarking Schemes}
Here, we employ the standard SD and CG methods proposed in \cite{John2006Chapter8} as benchmarks, where fixed step sizes are adopted instead of ILS. To facilitate performance comparison, these two benchmark algorithms should be adapted to our problem below, called penalty function and projection-driven normalized SD (PP-NSD) and penalty function and projection-driven normalized CG (PP-NCG), respectively.  

To mitigate potential instability arising from overly large or small gradient magnitudes under fixed-step updates, we normalize the update directions of CG or SD algorithms using their $l_2$ norm. Specifically, for the PP-NSD algorithm, the search direction is expressed by an $\epsilon_d$-regularized normalized gradient
\begin{align}
   {\bm d}_{i} &= -{\bm g}_{i}/(\|{\bm g}_{i}\|_2+\epsilon_d),\label{eq:direction NSD}
\end{align} 
where ${\bm g}_{i}={\bm \Theta}\nabla L({\bm \rho}_{i})$, and $\epsilon_d$ is a small positive constant to prevent division by zero.
Similarly, for the PP-NCG algorithm, the search direction is given by an $\epsilon_d$-regularized normalized CG, i.e.,
\begin{align} \small
   {\bm d}_{i+1} &= -({\bm g}_{i+1}+\varsigma_i{\bm d}_i)/(\|({\bm g}_{i+1}+\varsigma_i{\bm d}_i)\|_2+\epsilon_d),\label{eq:direction NCG}
\end{align} 
where the deflection factor $\varsigma_i$ is computed by \eqref{eq:deflection choice}.

The step size in both benchmark algorithms is given by 
\begin{align}
\upsilon_i = \min\{\epsilon_{\upsilon}\upsilon_i^{\rm b},\upsilon_0\},
\end{align}
with $\upsilon_i^{\rm b}$ given in \eqref{eq:step size range} and $\upsilon_0$ being a predetermined step size. Due to the uncertain convergence of the benchmarks, their maximum number of iterations is set to $I$. Moreover, due to the non-monotonic penalty function, the minimum may not occur at the final iteration but within a few steps before it. To address this, we introduce an alternative penalty stopping criterion given by $\mu_j\min\limits_{i+1-S\leq q\leq i+1}\{\alpha({\bm \rho}_{q})\}<\epsilon_{\mu}$ where $S$ represents the window size for searching the minimum penalty value. 
The detailed implementation steps of the PP-NSD algorithm are omitted here, as it can be easily constructed from the PP-NCG described in 	\textbf{Algorithm} \ref{alg:PP-NCG} by removing \textbf{step 4} and performing the direction update in \textbf{step 5} using \eqref{eq:direction NSD}.
\begin{algorithm}[!t]
   \small
	\renewcommand{\algorithmicrequire}{\textbf{Input:}}
	\renewcommand{\algorithmicensure}{\textbf{Output:}}
	\caption{The PP-NCG algorithm}
	\label{alg:PP-NCG}
   \textbf{Input:} Positive scalar termination factor $\epsilon_{\mu}$ and $\epsilon_{\rm th}$, a large integer $I$ and a small integer $S$, small positive constants $\sigma_{\upsilon}$, $\epsilon_L$, $\epsilon_d$ and $\epsilon_{\upsilon}$, initial step size $\upsilon_0$, initial penalty factor $\mu_1$ and its scaling factor $\varphi>1$; initial PA parameter ${\bm \rho}_0$ satisfying ${\bm 1}^{\rm T}{\bm \rho}_0=1$, projection operator ${\bm \Theta}$, ${\bm y}_1={\bm \rho}_1$, and ${\bm d}_1= -{\bm \Theta}\nabla L({\bm \rho}_1)/(\|{\bm \Theta}\nabla L({\bm \rho}_1)\|_2+\epsilon_d)$; a counter $k=0$ and $i=j=1$.\\
   \textbf{Output:} The PA vector ${\bm y}_j$.
	\begin{algorithmic}[1]
      \STATE Step size: $\upsilon_i = \min\{\epsilon_{\upsilon}\upsilon_i^{\rm b},\upsilon_0\}$. 
      \STATE PA update: ${\bm \rho}_{i+1}={\bm \rho}_{i}+\upsilon_i{\bm d}_{i}$.
      \STATE Convergence condition: If $|L({\bm \rho}_{i+1})-L({\bm \rho}_{i})|/|L({\bm \rho}_{i+1})|\leq\epsilon_{\rm th}$ or $i+1=I$, ${\bm y}_j={\bm \rho}_{i+1}$, go to \textbf{step 6}; Otherwise, go to next step.
      \STATE  Deflection choice: According to \eqref{eq:deflection choice}, if $\varsigma_i=\varsigma_i^{\rm HS}$, $k=k+1$, or $k=0$ otherwise; when $k=Q$, let $\varsigma_i=0$ and $k=0$; 
      \STATE Direction update: ${\bm d}_{i+1}$ through \eqref{eq:direction NCG}; let $i=i+1$, and go back to \textbf{step 1}.
      \STATE Penalty update: If $\mu_j\min\limits_{i+1-S\leq q\leq i+1}\{\alpha({\bm \rho}_{q})\}<\epsilon_{\mu}$, stop; Otherwise, let $\mu_{j+1}=\varphi\mu_j$, and ${\bm d}_{i+1} = -{\bm \Theta}\nabla L({\bm \rho}_{i+1})/(\|{\bm \Theta}\nabla L({\bm \rho}_{i+1})\|_2+\epsilon_d)$; let $j=j+1$ and $i=i+1$, and go to \textbf{step 1}.
	\end{algorithmic}
\end{algorithm}

   \vspace{-0.3em}
\section{Extension to Pure Sensing}\label{sec:Pure sensing}
In this section, we consider a pure sensing scenario with joint location and velocity estimation. The P-NCG-ILS algorithm is proposed for PA optimization in minimizing the total power under power budget constraints and CRLB constraints.

The constrained power minimization problem is given by 
\begin{subequations}  
   \begin{align}\small
      \arg\min\limits_{{\bm \rho}}&\ {\bm \rho}^{\rm T}{\bm 1},\label{eq:sensing problem}\\
      {\rm s.t.}\ & {\bm \rho}^{\rm T}{\bm 1}\leq1, \label{eq:total power limit} \\
         \ & \eqref{eq:individual power constraint}, \eqref{eq:location constraint}\ {\rm and}\ \eqref{eq:velocity constraint}.
   \end{align}  
\end{subequations}
Analogously to the derivations of PP-MCG-ILS, we set the penalty function for violating the inequality constraints as 
\begin{align} \small
   \alpha_{\rm s}({\bm \rho}) =\ & {f}_{\rm s}({\bm \rho})+\tilde{f}_l({\bm \rho})+\tilde{f}_v({\bm \rho})\notag\\
   &+\sum\nolimits_{n=1}^N\left(f_{\min,n}({\bm \rho})+f_{\max,n}({\bm \rho})\right),\label{eq:sensing penalty function}
\end{align}  
where $ {f}_{\rm s}({\bm \rho})$ denotes the penalty sub-function for the total power budget, i.e.,
\begin{align} \small
   {f}_{\rm s}({\bm \rho}) = (\max\{0,{\bm \rho}^{\rm T}{\bm 1}-1\})^q.
\end{align}
Then, the unconstrained optimization problem can be formulated as 
\begin{align} \small
   \arg\min\limits_{{\bm \rho}}\ L_{\rm s}({\bm \rho})={\bm \rho}^{\rm T}{\bm 1}+\mu\alpha_{\rm s}({\bm \rho}). \label{eq:sensing target with penalty}
\end{align}

Similarly to the PP-MCG-ILS algorithm, we also use the CG-based method to solve \eqref{eq:sensing target with penalty}. To avoid a potential numerical instability, we consider normalized CG search direction, given by ${\bm d}_{i+1}= \tilde{\bm d}_{i+1}/(\|\tilde{\bm d}_{i+1}\|_2+\epsilon_d)$ with $\tilde{\bm d}_{i+1}= -\nabla L_{\rm s}({\bm \rho}_{i+1})+\varsigma_i{\bm d}_i$, where the deflection computation method is the same as that of the PP-MCG-ILS algorithm. Additionally, ILS is used to determine the step size at each iteration.
The complete P-NCG-ILS algorithm is summarized in \textbf{Algorithm} \ref{alg:P-NCG}.
\begin{algorithm}[!t]
   \small
	\renewcommand{\algorithmicrequire}{\textbf{Input:}}
	\renewcommand{\algorithmicensure}{\textbf{Output:}}
	\caption{The P-NCG-ILS algorithm}
	\label{alg:P-NCG}
   \textbf{Input:} Positive scalar termination factors $\epsilon_{\mu}$ and $\epsilon_{\rm th}$, small positive constants $\sigma_{\upsilon}$, $\epsilon_L$ and $\epsilon_{\upsilon}$, initial step size $\upsilon_0$, initial penalty factor $\mu_1$ and its scaling factor $\varphi>1$; initial PA parameter ${\bm \rho}_1$ satisfying ${\bm 1}^{\rm T}{\bm \rho}_1=1$, ${\bm y}_1={\bm \rho}_1$, and ${\bm d}_1= -\nabla L_{\rm s}({\bm \rho}_1)/\|\nabla L_{\rm s}({\bm \rho}_1)\|_2$; a counter $k=0$ and $i=j=1$.\\
   \textbf{Output:} The PA vector ${\bm y}_j$.
	\begin{algorithmic}[1]
      \STATE Step size bound: Calculate $\upsilon_i^{\rm b}$ according to \eqref{eq:step size range}; 
      \STATE Step size: $\upsilon_i = \min\{\epsilon_{\upsilon}\upsilon_i^{\rm b},\upsilon_0\}$; while $L_{\rm s}({\bm \rho}_{i} + \upsilon_i{\bm d}_{i})-L_{\rm s}({\bm \rho}_{i})>\epsilon_L \upsilon_i{\bm d}_{i}^{\rm T}\nabla L({\bm \rho}_{i})$, let $\upsilon_i = \sigma_{\upsilon}\upsilon_i$. 
      \STATE PA update: ${\bm \rho}_{i+1}={\bm \rho}_{i}+\upsilon_i{\bm d}_{i}$.
      \STATE Convergence condition: If $|L_{\rm s}({\bm \rho}_{i+1})-L_{\rm s}({\bm \rho}_{i})|/|L_{\rm s}({\bm \rho}_{i+1})|\leq\epsilon_{\rm th}$, ${\bm y}_j={\bm \rho}_{i+1}$, and go to \textbf{step 5}; Otherwise, go to next step.
      \STATE Deflection choice: According to \eqref{eq:deflection choice}, if  $\varsigma_i=\varsigma_i^{\rm HS}$, $k=k+1$, or $k=0$ otherwise; when $k=Q$, let $\varsigma_i=0$ and $k=0$.
      \STATE Direction update: ${\bm d}_{i+1}= \tilde{\bm d}_{i+1}/(\|\tilde{\bm d}_{i+1}\|_2+\epsilon_d)$ with $\tilde{\bm d}_{i+1}= -\nabla L_{\rm s}({\bm \rho}_{i+1})+\varsigma_i{\bm d}_i$; let $i=i+1$, and go back to \textbf{step 1}.
      \STATE Penalty update: If $\mu_j\alpha({\bm y}_j)<\epsilon_{\mu}$, stop; Otherwise, $\mu_{j+1}=\varphi\mu_j$, and ${\bm d}_{i+1} = -\nabla L_{\rm s}({\bm \rho}_{i+1})$; let $j=j+1$ and $i=i+1$, and go to \textbf{step 1}.
	\end{algorithmic}
\end{algorithm}

\vspace{-0.5em}
\section{Algorithm Analysis}\label{sec:Algorithm analysis}
The proposed algorithms (e.g., PP-MCG-ILS and PP-MSD-ILS) can be readily extended to a broad class of nonlinear and nonconvex optimization problems. Such problems typically involve affine equality constraints, which, when present, are enforced via projection operators, while nonlinear or nonconvex constraints are handled through penalty functions. The ILS mechanism is further employed to improve convergence stability. As an example, the proposed P-NCG-ILS algorithm for the sensing-only scenario can be regarded as a direct extension of the PP-MCG-ILS algorithm.

In the following, we analyze the convergence behavior and computational complexity of the proposed algorithms.

\vspace{-1em}
\subsection{Convergence Analysis}
We now provide the convergence analysis of the proposed PP-MCG-ILS and PP-MSD-ILS algorithms. Before delving to the details, we define the increment of the parameter update and the objective function respectively as
\begin{align} \small
   \Delta {\bm \rho}_{i+1} &\triangleq  {\bm \rho}_{i+1}-{\bm \rho}_i = v_i {\bm d}_i,
\end{align}
and 
\begin{align}
   \Delta L({\bm \rho}_{i+1}) \triangleq L({\bm \rho}_{i+1})-L({\bm \rho}_{i}).
\end{align} 

\textbf{Lemma} \textbf{1}. \emph{The sufficient and necessary condition for $\|\Delta {\bm \rho}_{i+1}\|\rightarrow 0$ and $\Delta L({\bm \rho}_{i+1})\rightarrow 0$ is that, there exists an integer $I$ such that, ${\bm d}_{i}\rightarrow {\bm 0}$ or $v_i\rightarrow 0$ when $i>I$.}

\emph{Proof.} The proof is given in Appendix \ref{sec: proof of lemma 1}. \hfill$\blacksquare$

{\textbf{Definition.}} \emph{A function $L({\bm \rho})$ is Lipschitz gradient continuous over a set $\Omega$, if there exists a constant $C>0$, $\|\nabla L({\bm \rho}_i)-\nabla L({\bm \rho}_j)\|\leq C\|{\bm \rho}_i-{\bm \rho}_j\|$ for any ${\bm \rho}_i, {\bm \rho}_j\in \Omega$ \cite{Micheal2006}.}


\textbf{Theorem} \textbf{1}. \emph{Consider the CRLB-based penalty objective $L(\boldsymbol{\rho})$ defined in \eqref{eq:target with penalty},
over a nonempty and bounded convex feasible set $\Omega \subset \mathbb{R}^n$.
If $L$ is continuously differentiable and admits a Lipschitz continuous gradient on $\Omega$,
the proposed ILS-based schemes (PP-MCG-ILS and PP-MSD-ILS) generate a sequence
$\{\boldsymbol{\rho}_i\}$ whose limit points are first-order stationary with a zero descent direction ${\bm d}_{i}\rightarrow {\bm 0}$.
}

\textbf{Theorem} \textbf{1} is included to guarantee the existence of a well-defined and bounded step size
for the proposed ILS-based PA algorithms. 

\emph{Proof.} The proof is given in Appendix \ref{sec: proof of theorem 1}. \hfill$\blacksquare$


We now establish the Lipschitz continuity of the gradients of the objective functions in \eqref{eq:unconstrained optimization} and \eqref{eq:sensing target with penalty} by showing that their second-order derivatives are uniformly bounded over the feasible set.

Firstly, since $\Omega \subset \mathbb{R}^N $ is compact and all CRLB-related matrices, e.g., $ {\bm P}(\rho) $, $ {\bm Y}(\rho) $, and their inverses depend smoothly on ${\bm \rho}$,
their spectral norms and those of their derivatives are uniformly bounded over $\Omega$.
As a result, the operator norm of the Hessian of $f_l({\bm \rho})$ is uniformly bounded.

Moreover, we know that ${\bm C}_L({\bm \rho})$ is uniformly positive definite. It follows that ${\bm C}_L({\bm \rho})$ is continuously differentiable, and so is ${\rm tr}\{{\bm C}_L({\bm \rho})\}$.
By the chain rule, we have
\begin{equation} 
\nabla \tilde{f}_l({\bm \rho}) = 
\begin{cases}
2 f_l({\bm \rho}) \nabla f_l({\bm \rho}), & \text{if } f_l({\bm \rho}) > 0 \\
0, & \text{otherwise}.
\end{cases}
\end{equation}
where $\nabla f_l({\bm \rho}) =-{\rm tr}\{{\bm C}_L({\bm \rho})\nabla ({\bm P}-{\bm V}{\bm Y}^{-1}{\bm V}^{\rm T}) {\bm C}_L({\bm \rho})\}$. Since the matrix inverse and trace are smooth on the set of positive definite matrices, $\nabla f_l({\bm \rho})$ is continuously differentiable. Together with the uniform boundedness of the Hessian $\nabla^2 f_l({\bm \rho})$ over $\Omega$,
this implies that $\nabla f_l({\bm \rho})$ is Lipschitz continuous over $\Omega$.
Consequently, $ \nabla \tilde{f}_l({\bm \rho}) $ is Lipschitz continuous. A similar argument applies to $ \nabla \tilde{f}_v({\bm \rho}) $. 

Furthermore, $f_{\min,n}({\bm \rho})$, $ f_{\max,n}({\bm \rho}) $ and ${f}_{\rm s}({\bm \rho})$ are piecewise quadratic with finitely many breakpoints, and their gradients are globally Lipschitz continuous over $\Omega$.
Hence, $ \nabla_{\bm \rho} \alpha({\bm \rho}) $ and $ \nabla_{\bm \rho} \alpha_{\rm s}({\bm \rho}) $ are Lipschitz continuous over $ \Omega $, and the same applies to $ \nabla_{\bm \rho} L({\bm \rho}) $ and $ \nabla_{\bm \rho} L_{\rm s}({\bm \rho}) $.

Finally, since $ {\bm \Theta} $ is a constant matrix with $ \|{\bm \Theta}\|_2 = 1 $, and $ \nabla_{\bm \rho} f({\bm \rho}) $ is Lipschitz continuous, the composition $\nabla\mathcal{L} ({\bm \varpi})={\bm \Theta}\nabla L({\bm \rho})$ is also Lipschitz continuous. In other words, there exists a constant $ C > 0 $, independent of ${\bm \varpi}$, such that
\begin{equation}
\small
\|\nabla_{{\bm \varpi}} \mathcal{L}({\bm \varpi}_1) - \nabla_{{\bm \varpi}} \mathcal{L}({\bm \varpi}_2)\|_2 \le C \|{\bm \varpi}_1 - {\bm \varpi}_2\|_2, \forall {\bm \varpi}_1, {\bm \varpi}_2 \in \Omega_{\bm \varpi} \notag
\end{equation}  
Therefore, $ \mathcal{L}({\bm \varpi}) $ is Lipschitz gradient continuous on $ \Omega_{\bm \varpi} $.
\hfill$\blacksquare$

   \vspace{-1em}
\subsection{Complexity Analysis}
We now qualitatively analyze and compare the computational complexity of the proposed algorithms and their benchamrks.

We observe that the increased computations of the PP-MCG-ILS algorithm, compared to the PP-MSD-ILS algorithm, primarily stems from the computation of $\varsigma_i^{\rm b}$ and $\varsigma_i^{\rm HS}$ at each iteration, which involves approximately $4N$ multiplications and 2 divisions. In both algorithms, the dominant computational burden lies in \emph{evaluating the objective function during the ILS} and \emph{computing the gradients for direction updates}, which are inherently tied to the CRLB expressions. Therefore, a more meaningful comparison of the overall computational cost focuses on the number of ILS steps performed per iteration. This number is influenced by the step size threshold, given by the right-hand side of \eqref{eq:step size threshold}, the initial step size $\upsilon_{i,0}$ in \eqref{eq:step size initial}, and the scaling factor. At a given iterate ${\bm \rho}_i$, a larger search direction norm $\|{\bm d}_i\|$ generally leads to a smaller $\upsilon_{i,0}$ and a reduced step size threshold. As a consequence, the number of ILS steps per iteration is expected to remain comparable between the PP-MCG-ILS and PP-MSD-ILS algorithms.
Therefore, their relative performance is primarily characterized by their convergence rates and the quality of the converged solutions.

We next compare algorithms with and without ILS. As discussed above, the number of ILS steps directly affects the number of objective-function evaluations. Accordingly, benchmark algorithms without ILS incur a lower step size selection cost per iteration compared with the PP-MCG-ILS and PP-MSD-ILS algorithms. However, all four algorithms exhibit similar computational complexity in the direction update, since gradient evaluations dominate this stage.

Finally, the overall computational complexity is strongly influenced by the convergence rate, or equivalently, by the total number of iterations required to reach convergence. As demonstrated in the simulation results, the proposed PP-MCG-ILS algorithm converges significantly faster than the benchmark methods, both with and without ILS. This faster convergence substantially reduces the overall computational effort, particularly by alleviating the repeated evaluation of computationally intensive gradients.

   \vspace{-0.5em}
\section{Simulation Results}\label{sec:simulations}
In this section, we conduct numerical simulations to evaluate the proposed PP-MCG-ILS, PP-MSD-ILS and P-NCG-ILS algorithms for PA optimization. All APs are assumed to be stationary.
We employ the orthogonal chirp division multiplexing (OCDM) \cite{Ouyang2016} waveform with a single Gaussian pulse shaping. The carrier frequency is $3$GHz. The lowpass equivalent is formulated by
\begin{align}
s_n(t) = (2/T^2)^{1/4}e^{-\pi t^2/T^2}e^{j\pi M/T^2(t-(n-1)T/M)^2},
\end{align}
where $M$ is the number of OCDM chirps, and $T$ is proportional to the effective pulse time width. 

   \vspace{-1em}
\subsection{CRLB Validation}
To verify the approximate CRLB, we first compare it with the mean squared error (MSE) of MLE results across the considered SCNR range. To maintain consistency with existing works \cite{Godrich2010, Godrich2011, Ai2015}, we use the signal energy-to-noise ratio (SENR), defined as  $\tilde\delta_{\rm cn}\triangleq P/\sigma_{\rm cn}^2$, instead of the SCNR $\delta_{\rm cn}$ in evaluating the CRLBs. Due to the high complexity of four-dimensional MLE, we consider a simple $4\times 3$ nonsymmetric multi-static radar configuration given in Fig. \ref{fig:Location Radar}. Without loss of generality, we consider the OCDM waveform with $M=16$, $T=10^{-2}$ s and a sampling rate of $f_s=1$ kHz. 



\begin{figure}[!t]
   \centerline
   {\includegraphics[width=0.3\textwidth]{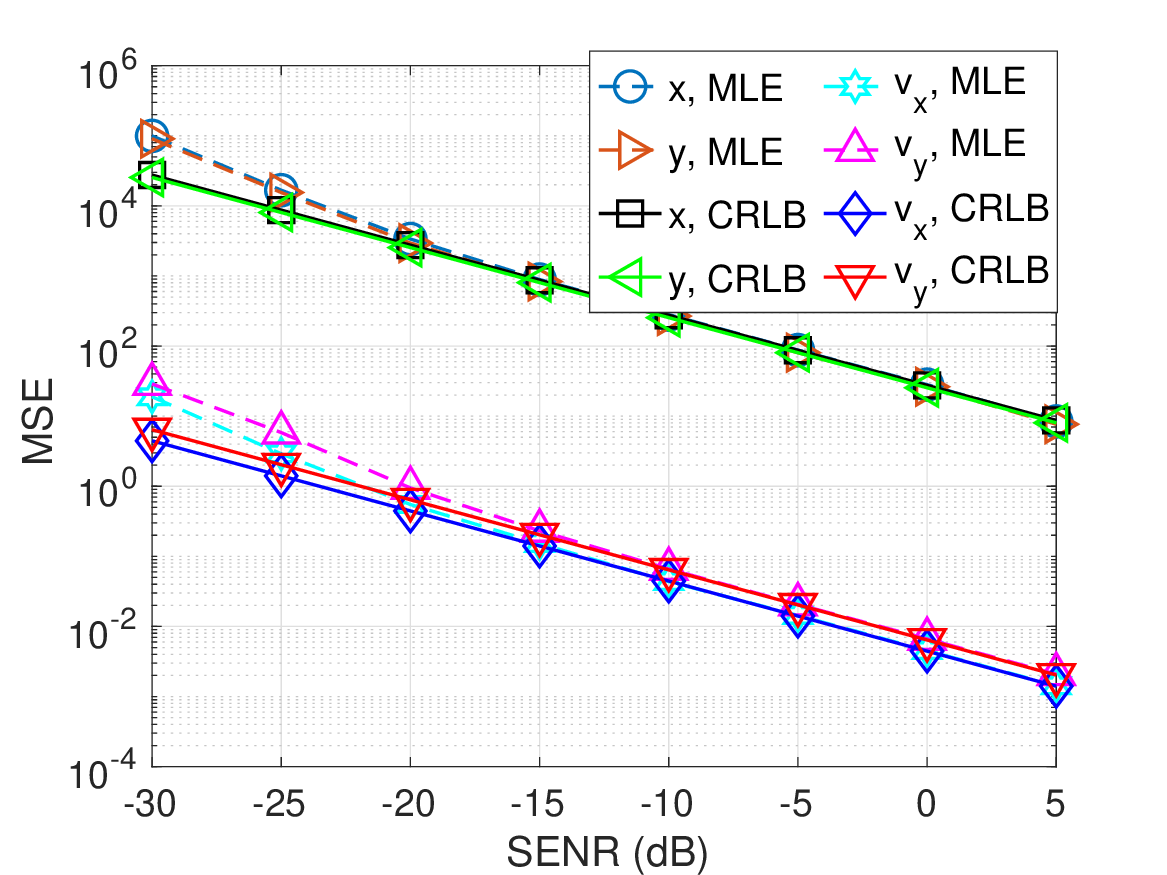}}
   \caption{The MSE versus the CRLB under different SENRs.}\label{fig:MSE CRLB}
   \vspace{-0.3cm}
\end{figure}

Fig. \ref{fig:MSE CRLB} compares the MSE of the MLE with the derived CRLB. It is observed that the derived CRLBs closely match the MSEs for SENR values above $-20$ dB. The small discrepancies in the low-SENR regime are attributed to the suboptimal behavior of the likelihood function under noise-dominated conditions. Overall, these results validate the effectiveness and accuracy of the proposed CRLBs.

   \vspace{-1em}
\subsection{Power Allocation for ISAC}
\begin{figure}[!t] 
   \centering
   \subfigure[]{\label{fig:Location Radar}\includegraphics[width=0.24\textwidth]{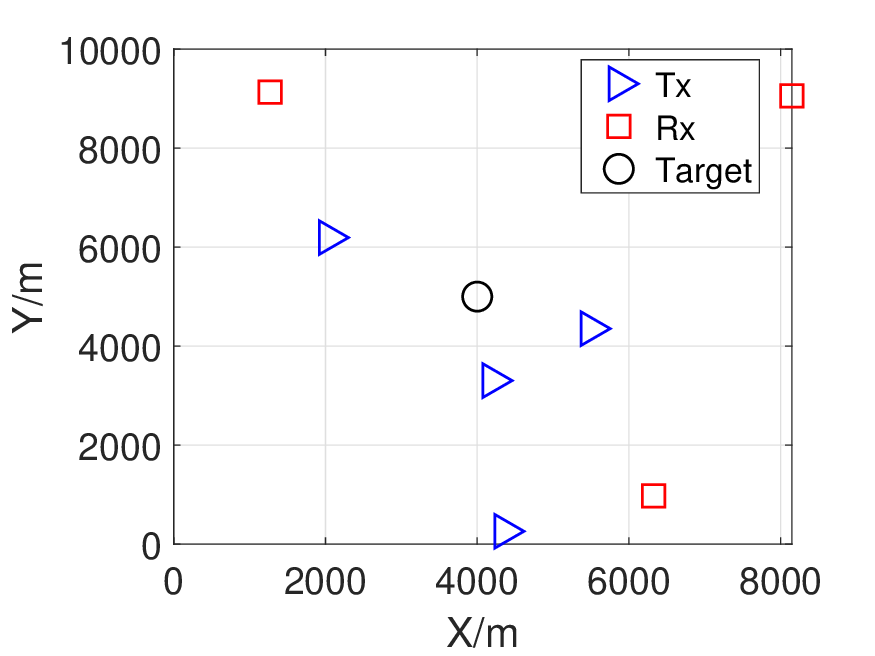}}
   \subfigure[]{\label{fig:Location CF-MIMO-ISAC}\includegraphics[width=0.24\textwidth]{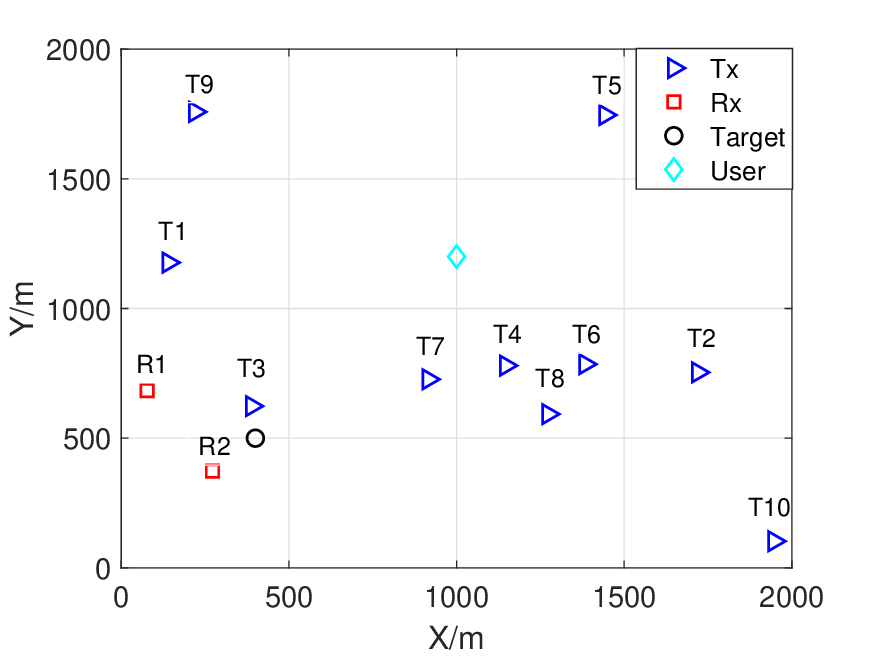}}
   \caption{The AP location distribution of the ISAC system: (a) $4\times 3$ radar setting; (b) CF MIMO-ISAC setting.}
   \label{fig:Location ISAC}
   \vspace{-0.3cm}
\end{figure}

The performance of the proposed PA algorithms is evaluated by comparison with the PP-NCG and PP-NSD benchmark algorithms. To ensure communication quality, a larger number of APs are designated for transmitting ISAC signals, while a smaller subset is used for receiving target-reflected signals. Without loss of generality, we consider a CF MIMO-ISAC system comprising 10 transmitters and 2 receivers in Fig. \ref{fig:Location CF-MIMO-ISAC}.

To guarantee the CRLB effectiveness, we assume the minimum PA coefficient vector is ${\bm \rho}_{\min}=0.01\times{\bm 1}_N$. The PA budget is ${\bm \rho}_{\max}=0.3\times{\bm 1}_N$. The SENR is set to be $-10$ dB. 
Within the coherence time for RCS, a typical modulus-squared RCS matrix is given by {\footnotesize{$\begin{bmatrix}
0.37&0.70&1.38&0.65& 2.40& 0.18& 0.25& 0.82& 0.14& 0.35	\\ 0.40\!&\! 2.05& 1.14& 0.42& 0.03& 0.11& 2.02& 1.65& 1.24& 3.65\end{bmatrix}^{\rm T}$}}. Similarly, the channel gains between the user and the transmitters are assumed to be {\small{${\bm g}=[2.11, 12.57, 5.63, 0.75, 0.61, 1.75, 0.20, 2.34, 14.79, 9.68]^{\rm T}$}}. We consider the OCDM waveform with $M=16$, $T=10^{-2}$ s and a sampling rate of $f_s=10$ kHz. Under the given ISAC system setup and waveform configuration, the CRLB thresholds for position and velocity are set to $[250,0.13]$. 

The PA vector is initialized as the uniform PA vector ${\bm \rho_0}={\bm 1}_N/N$, unless noted otherwise. The initial penalty factor is set to be $\mu_1=10^4$. We hereafter consider the ILS controlling factor $\epsilon_L=1e-3$, the initial step size controlling factor $\epsilon_{\upsilon} =0.9$, the penalty factor scaling coefficient $\phi=10$, and the penalty threshold $\epsilon_{\mu}=10^{-3}$, unless noted otherwise. The other simulation conditions are given in Table \ref{tab:parameter}. 

\begin{table*}[!t]
       \centering
       \small
       \caption{Simulation parameters}
       \vspace{-0.1cm}
      \label{tab:parameter}	
       \begin{tabular}{l|cccc}
           \Xhline{2pt}
           Algorithms&PP-MCG-ILS&PP-MSD-ILS&PP-NCG&PP-NSD\\\Xhline{1pt}
           Fig. \ref{fig:SINR_sequence}&$\{\epsilon_{\rm th},\sigma_{\upsilon}\}=\{1e-11,0.5\}$&same as PP-MCG-ILS&$\upsilon\in\{4e-5,2e-5\}$&same as PP-NCG\\
           Figs. \ref{fig:PA coefficient} and \ref{fig:CRLB sum}&same as Fig. \ref{fig:SINR_sequence}&---&$\upsilon=2e-5$&---\\
           Figs. \ref{fig:SINR} and \ref{fig:SINR_gain}&$\epsilon_{\rm th}=1e-12$, $\sigma_{\upsilon}=0.7$&same as PP-MCG-ILS&$\upsilon=2e-5$&same as PP-NCG\\
           Figs. \ref{fig:Location_threshold} and \ref{fig:Velocity_threshold}&$\{\epsilon_{\rm th},\sigma_{\upsilon}\}=\{1e-11,0.7\}$&---&---&---\\
           \Xhline{2pt}
      \end{tabular}
   \vspace{-0.5cm}
\end{table*}
\begin{figure}[!t]
   \centerline
   {\includegraphics[width=0.35\textwidth]{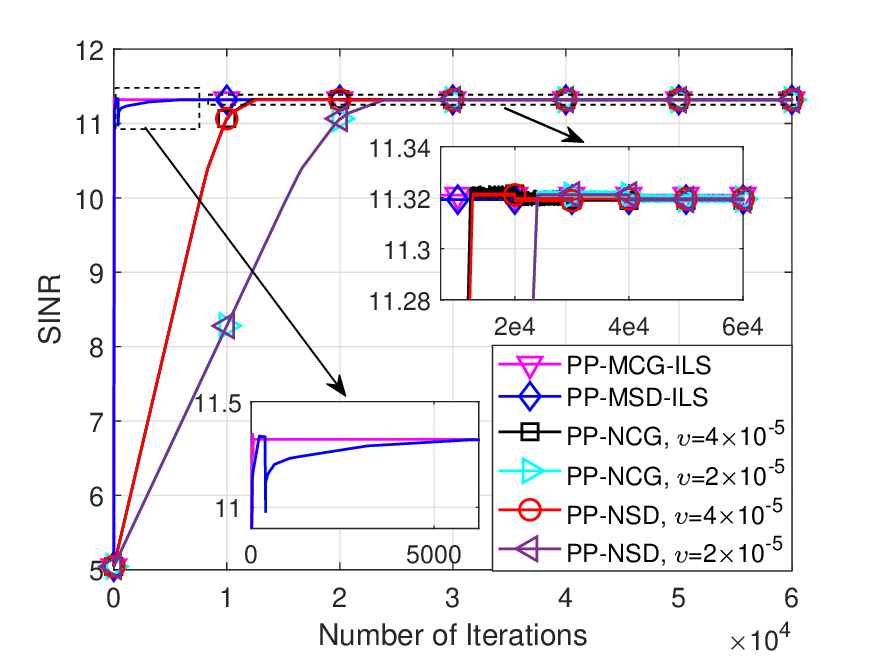}}
   \caption{SINR convergence behavior.}\label{fig:SINR_sequence}
   \vspace{-0.5cm}
\end{figure}
\begin{figure}[!t] 
   \centering
   \subfigure[]{\label{fig:rho_MCG_ILS}\includegraphics[width=0.24\textwidth]{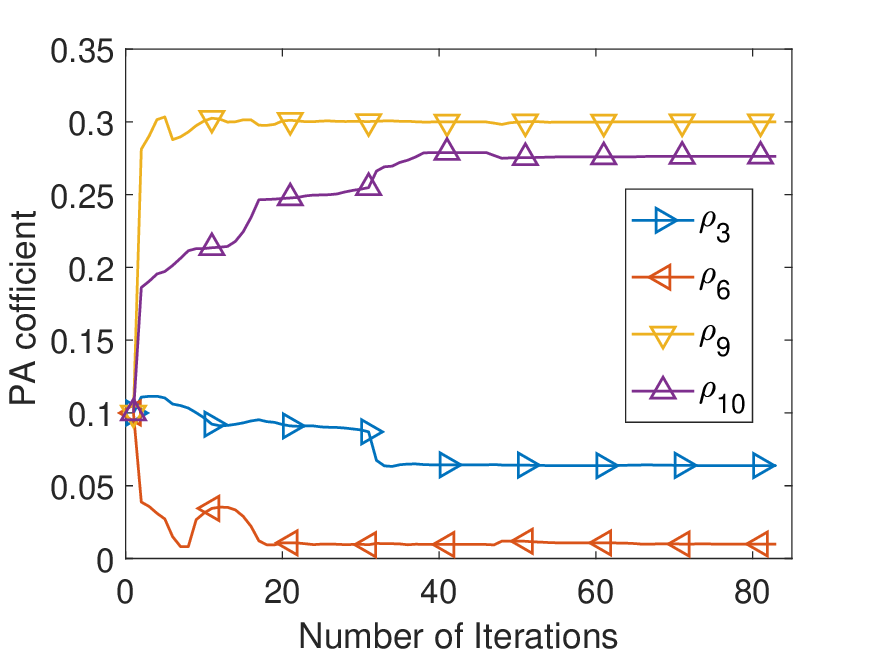}}
   \subfigure[]{\label{fig:rho_NCG}\includegraphics[width=0.24\textwidth]{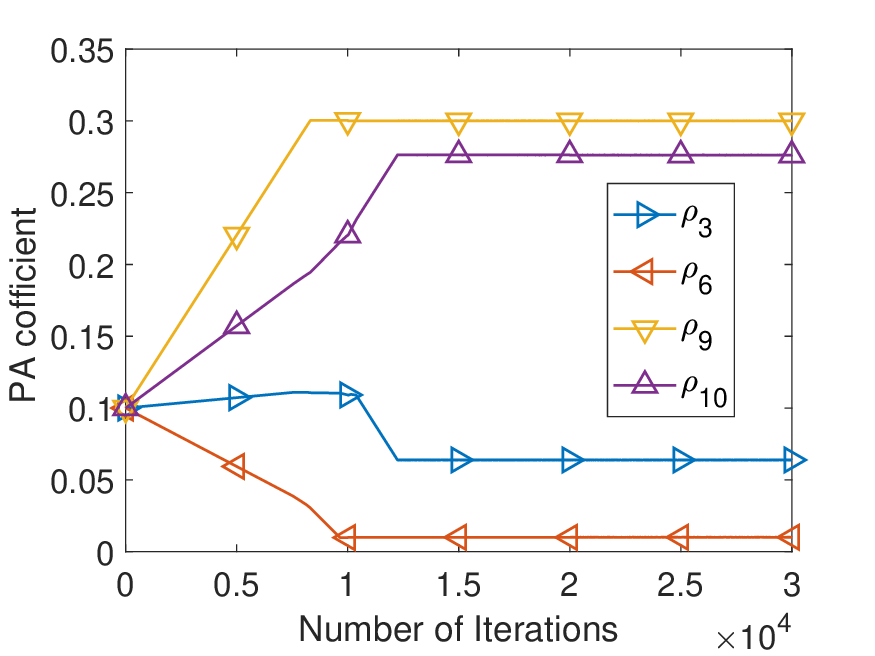}}
   \caption{PA: (a) PP-MCG-ILS; (b) PP-NCG with $\upsilon=2\times10^{-5}$.}
   \label{fig:PA coefficient}
   \vspace{-0.5cm}
\end{figure}
\begin{figure}[!t] 
   \centering
   \subfigure[]{\label{fig:CRLB_MCG_ILS}\includegraphics[width=0.24\textwidth]{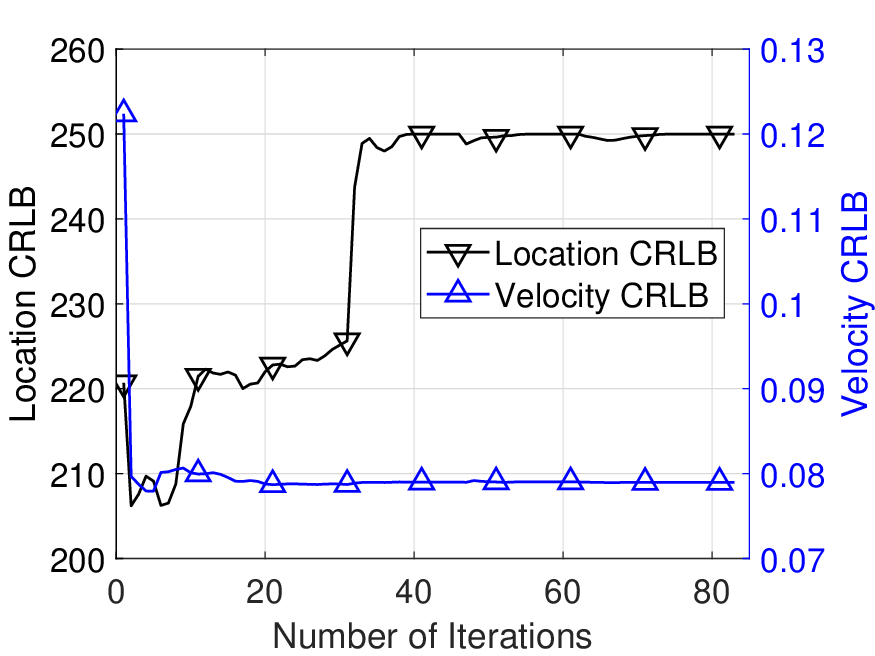}}
   \subfigure[]{\label{fig:CRLB_NCG}\includegraphics[width=0.24\textwidth]{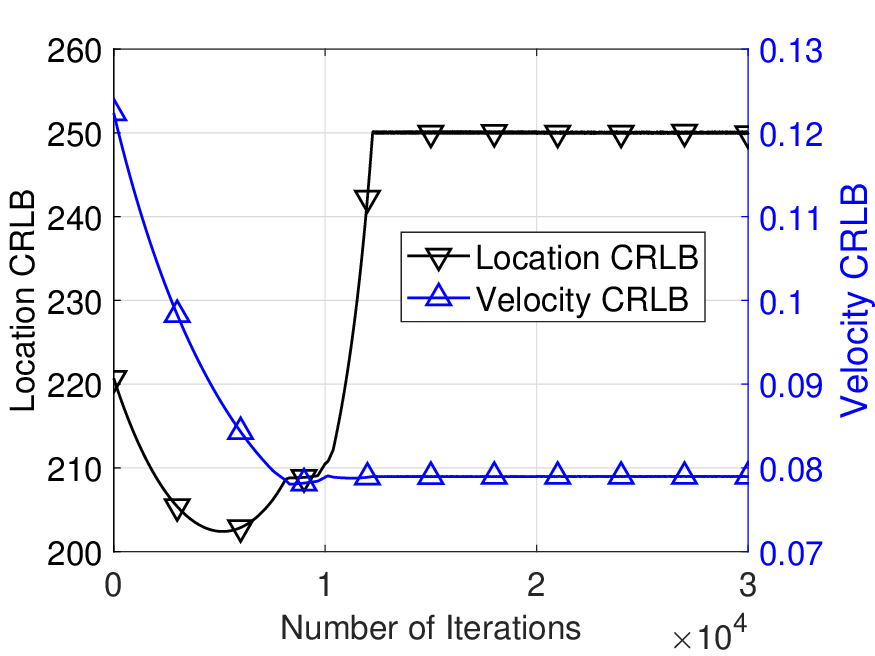}}
   \caption{CRLBs: (a) PP-MCG-ILS; (b) PP-NCG with $\upsilon=2\times10^{-5}$.}
   \label{fig:CRLB sum}
   \vspace{-0.5cm}
\end{figure}

We begin by evaluating the convergence and steady-state performance of the proposed algorithms. The SINR results corresponding to the iterative PA vectors for all four algorithms are depicted in Fig.~\ref{fig:SINR_sequence}. As observed, the PP-MCG-ILS algorithm exhibits the fastest convergence, followed by the PP-MSD-ILS. In contrast, the benchmark algorithms converge significantly more slowly. Furthermore, the benchmarks without ILS display noticeable numerical fluctuations, particularly under large fixed step sizes, suggesting potential instability and the risk of divergence. Therefore, we refrain from further increasing the step sizes in these simulations.

The evolutions of the PA coefficients and the corresponding CRLB values over iterations are illustrated in Fig. \ref{fig:PA coefficient} and Fig. \ref{fig:CRLB sum}, respectively. Overall, the results demonstrate that the PP-MCG-ILS algorithm achieves significantly faster convergence than the PP-NCG algorithm. As shown in Fig. \ref{fig:PA coefficient}, all algorithms eventually converge to the same PA vector, given by ${\bm \rho} = [0.01,\ 0.30,\ 0.0638,\ 0.01,\ \dots,\ 0.01,\ 0.30,\ 0.2762]^{\rm T}$, which satisfies the power constraints. Notably, transmitters 2, 9, and 11 are allocated more power due to their superior communication channel gains. Furthermore, Fig. \ref{fig:CRLB sum} dipicts the exploration process of the CRLB values under threshold-based constraints, in particular for the location CRLB, thereby validating the effectiveness of the proposed optimization framework.  

\begin{figure}[!t]
   \centerline
   {\includegraphics[width=0.35\textwidth]{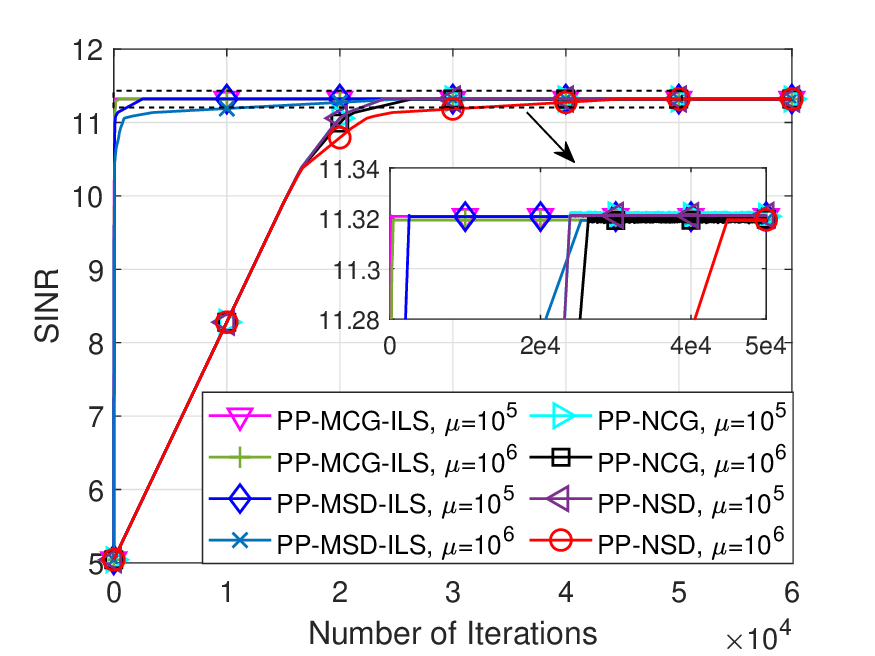}}
   \caption{SINR with initial uniform PA}\label{fig:SINR}
   \vspace{-0.5cm}
\end{figure}

\begin{figure}[!t]
   \centerline
   {\includegraphics[width=0.35\textwidth]{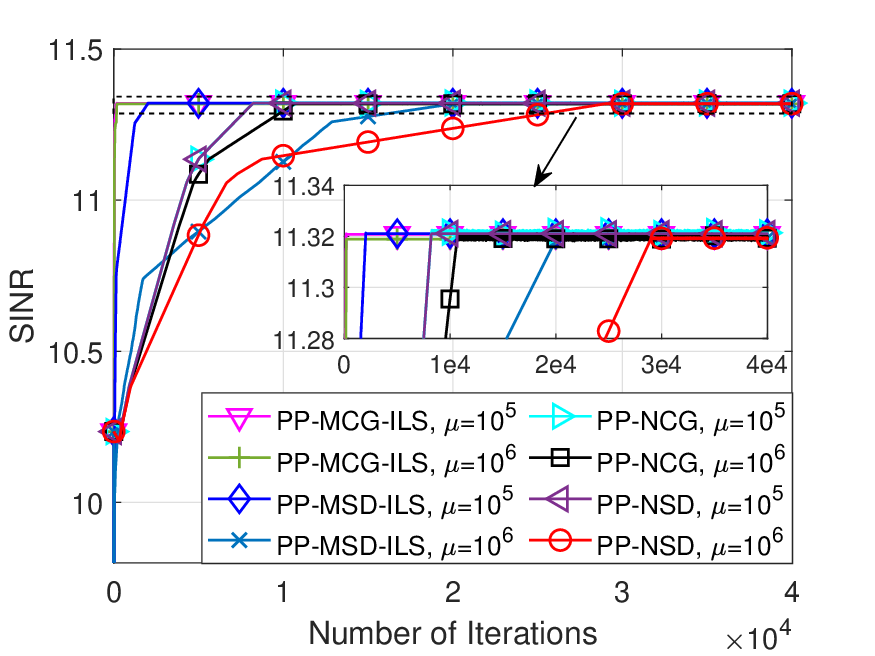}}
   \caption{SINR with initial heuristic PA}\label{fig:SINR_gain}
   \vspace{-0.5cm}
\end{figure}

The sequential penalty-based algorithms use the optimal solution from the previous loop as the initialization for the next loop with an updated penalty factor. To further examine the impact of initialization, we fix the penalty factor and compare the ISAC performance under different PA initialization strategies. In addition to uniform initialization, a heuristic scheme is considered, where the initial PA is set proportional to the communication channel gain, i.e., ${\bm \rho}_0 = {\bm g}/\sum_n |g_n|^2$.
The results in Figs.~\ref{fig:SINR} and~\ref{fig:SINR_gain} show that a smaller penalty factor ($\mu = 10^{5}$) yields faster convergence than $\mu = 10^{6}$, especially for the benchmark algorithms. Moreover, heuristic PA initialization significantly accelerates convergence for all methods compared to uniform initialization.
When comparing the fixed-penalty schemes with uniform initialization (Fig.~\ref{fig:SINR}) to the sequential penalty-based schemes (Fig.~\ref{fig:SINR_sequence}), the PP-MCG-ILS and PP-NCG algorithms exhibit similar convergence behavior, whereas the PP-MSD-ILS and PP-NSD algorithms converge more slowly. With heuristic initialization (Fig.~\ref{fig:SINR_gain}), PP-NCG and PP-NSD achieve noticeable performance gains, while PP-MCG-ILS maintains robust and superior convergence and PP-MSD-ILS shows a slight degradation.

To summarize, while the benchmark algorithms can benefit from specific conditions, such as an empirically tuned penalty factor and a well-initialized starting point, their performance is less robust and generalizable. In contrast, the PP-MCG-ILS algorithm exhibits superior convergence behavior, consistently outperforming the benchmark methods, irrespective of initialization or penalty factor selection. 

\begin{figure}[!t] 
   \centering
   \subfigure[]{\label{fig:tradeoff_SINR_CRLB}\includegraphics[width=0.24\textwidth]{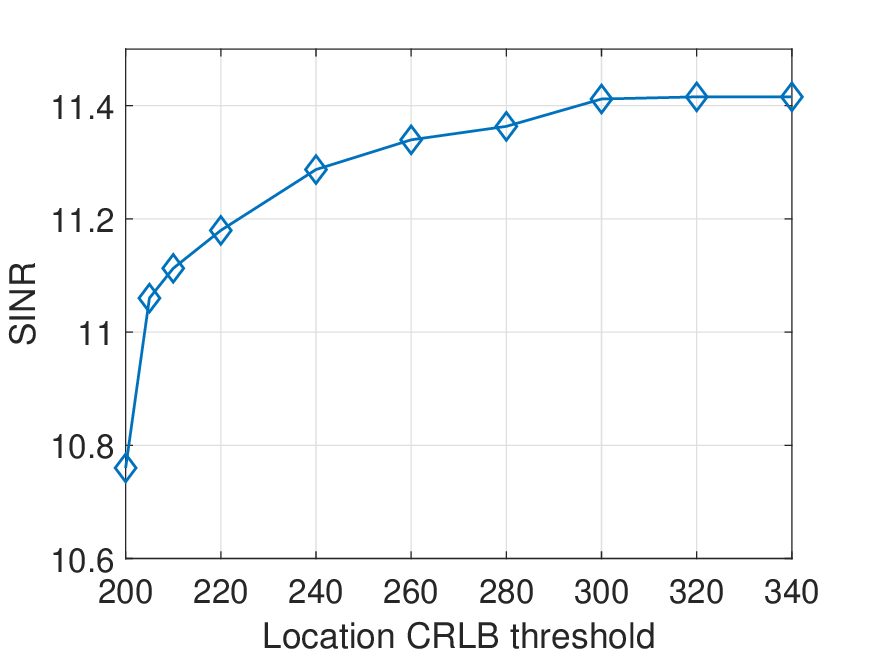}}
   \subfigure[]{\label{fig:Location_threshold_CRLB}\includegraphics[width=0.24\textwidth]{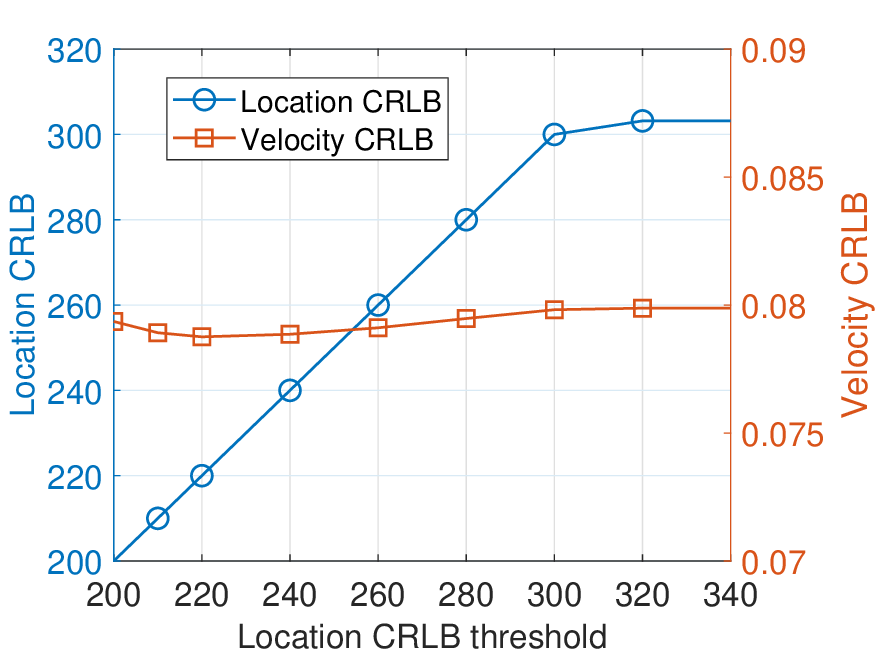}}
   \caption{Steady-state SINRs and CRLBs: (a) SINR; (b) CRLB.}
   \label{fig:Location_threshold}
   \vspace{-0.5cm}
\end{figure}

We now investigate the impact of the CRLB thresholds on the performance of the proposed PP-MCG-ILS algorithm. As shown in Fig.~\ref{fig:tradeoff_SINR_CRLB}, relaxing the location CRLB threshold results in a gradual increase in the steady-state SINR, which eventually saturates. This behavior is consistent with the trend observed in Fig.~\ref{fig:Location_threshold_CRLB}, where the achieved location CRLB closely approaches its prescribed threshold and then converges. These results illustrate the inherent trade-off between communication SINR and location CRLB.
\begin{figure}[!t] 
   \centering
   \subfigure[]{\label{fig:tradeoff_SINR_CRLB_v}\includegraphics[width=0.24\textwidth]{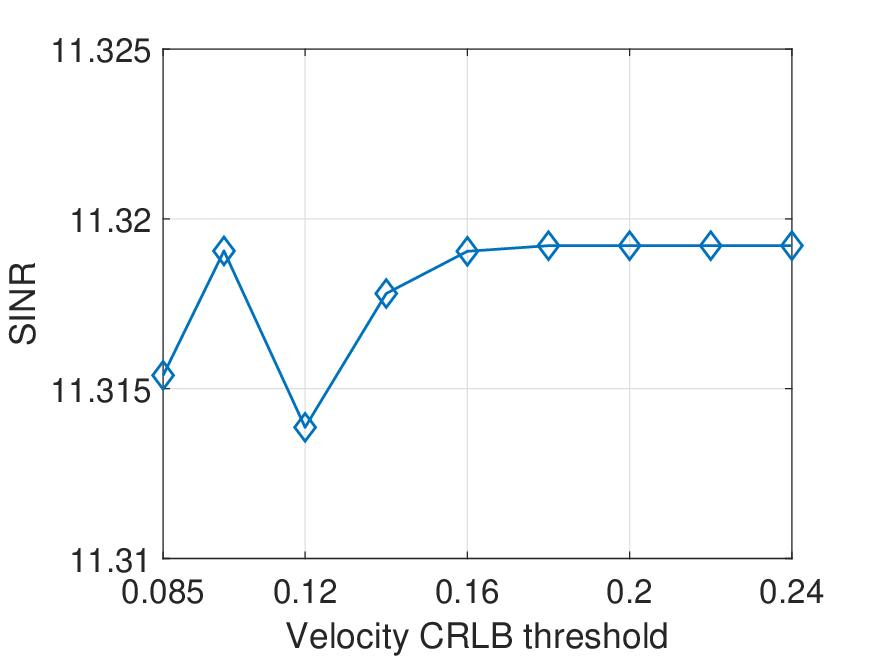}}
   \subfigure[]{\label{fig:Velocity_threshold_CRLB}\includegraphics[width=0.24\textwidth]{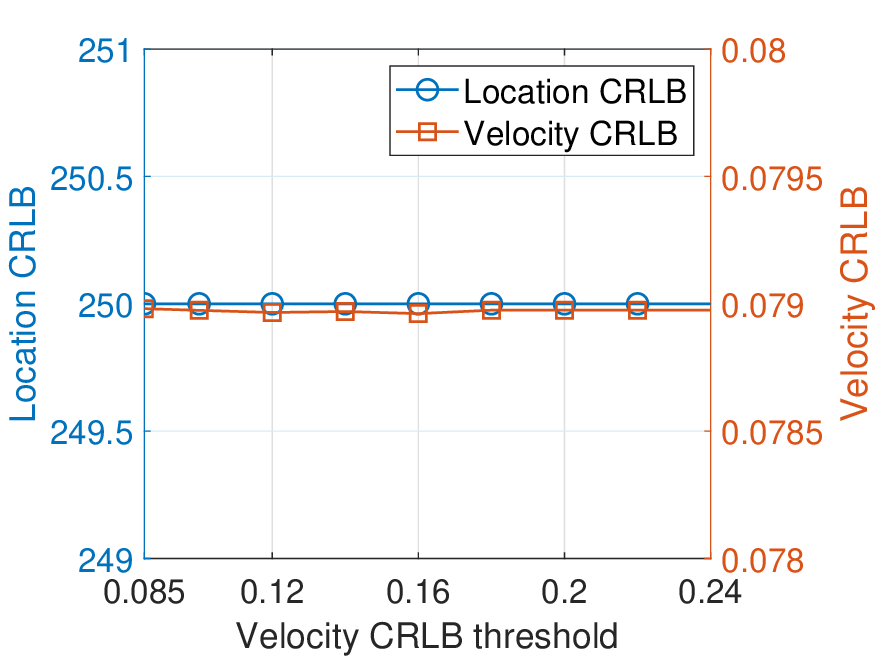}}
   \caption{Steady-state SINRs and CRLBs: (a) SINR; (b) CRLB.}
   \label{fig:Velocity_threshold}
   \vspace{-0.5cm}
\end{figure}

However, the velocity CRLB in Fig.~\ref{fig:Location_threshold_CRLB} is largely insensitive to variations in the location CRLB threshold. Similarly, Fig.~\ref{fig:Velocity_threshold} shows that changing the velocity CRLB threshold has a negligible impact on the performance of the PP-MCG-ILS algorithm. The observed saturation and difference of the CRLB with respect to its threshold is likely a consequence of the nonlinear structure of the CRLB formulation and the compact feasible region imposed by the power constraints.

\begin{figure}[!t] 
   \centering
   \subfigure[]{\label{fig:total power}\includegraphics[width=0.24\textwidth]{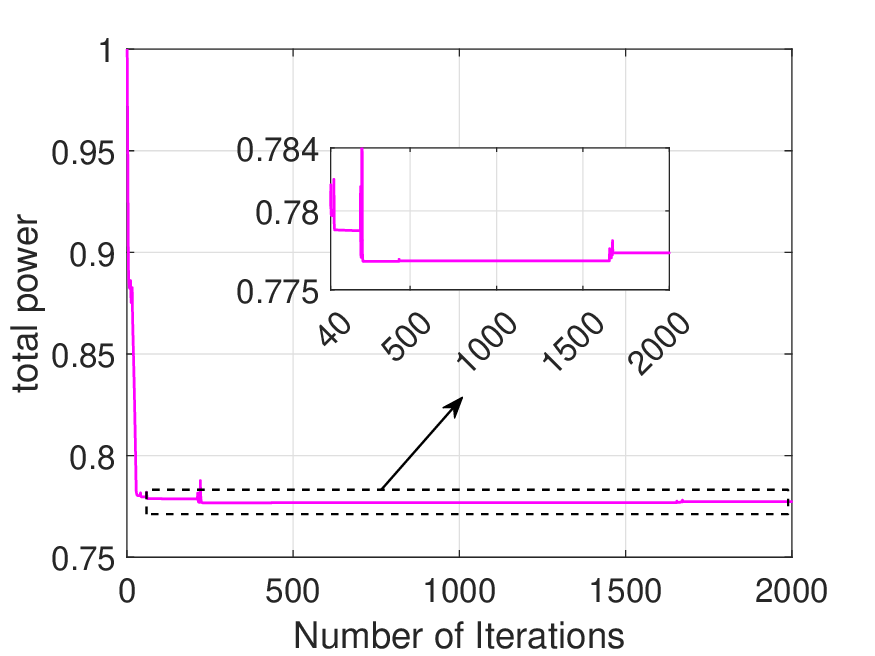}}
   \subfigure[]{\label{fig:CRLB_sensing}\includegraphics[width=0.24\textwidth]{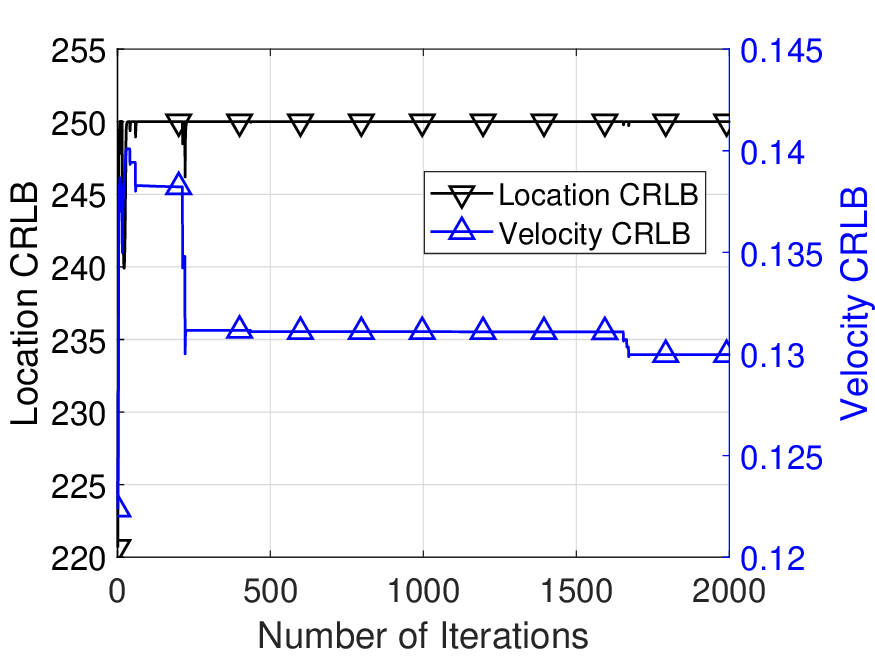}}
   \caption{Learning results for pure sensing: (a) total power; (b) CRLBs.}
   \label{fig:sensing}
   \vspace{-0.5cm}
\end{figure}

   \vspace{-1em}
\subsection{Power Allocation for Pure Sensing}
We now evaluate the proposed P-MCG-ILS algorithm on total power minimization in pure sensing scenario. The initial penalty factor is set to be $\mu_1=1$, $\sigma_{\upsilon}=0.2$ and the penalty threshold $\epsilon_{\mu}=10^{-6}$. The other simulation conditions are the same as the PP-MCG-ILS algorithm in Fig. \ref{fig:SINR_sequence}.    

The iterative results for total power optimization and the corresponding CRLB performance are illustrated in Fig. \ref{fig:sensing}. As seen in Fig. \ref{fig:total power}, the algorithm demonstrates rapid convergence. Under the same CRLB threshold settings, the total power consumption in the pure sensing scenario is significantly lower than that in the ISAC case. This is expected, as ISAC systems must allocate sufficient power to meet both sensing and communication requirements, particularly for maximizing communication performance.

Moreover, both the location and velocity CRLBs in the pure sensing case closely align with their respective thresholds. In contrast, in the ISAC scenario (refer to Fig. \ref{fig:CRLB_MCG_ILS} and Fig. \ref{fig:Velocity_threshold_CRLB}), the velocity CRLB matches its threshold loosely. This difference arises because, in pure sensing, any enhancement in CRLB performance directly benefits from increased sensing power. However, in ISAC, PA is optimized primarily for communication performance, with CRLB constraints being satisfied, potentially loosely, within acceptable margins.

\vspace{-0.2em}
\section{Conclusions}\label{sec:conclusion}
This paper presented a comprehensive investigation of the PA problem in a CF MIMO-ISAC framework, where joint location and velocity estimation were considered as sensing tasks. Newly derived approximate CRLBs were employed as sensing performance metrics, together with the user receive SINR as the communication metric. To address the inherent nonlinearity and nonconvexity of the resulting optimization problem, the PP-MCG-ILS algorithm was proposed. In addition, the P-NCG-ILS algorithm was developed to handle the power minimization problem in pure sensing scenarios. We further conducted a detailed convergence analysis and provided a qualitative comparison of the computational complexity of the proposed methods. Simulation results confirmed the accuracy of the derived CRLBs and demonstrated the effectiveness of the proposed PA strategies in enhancing both pure sensing and overall ISAC performance. These findings provide valuable insights into resource-efficient design for next-generation CF MIMO-ISAC networks and lay the groundwork for future extensions involving more complex sensing tasks and dynamic resource management.
\vspace{-0.11em}
\appendices
\section{Relevant parameters of the derived CRLBs}\label{sec: CRLB details}
The waveform-related parameters $a_{n,k}$, $b_{n,k}$ and $d_{n,k}$, are given by
\begin{equation}
\small
[a_{n,k},\ b_{n,k},\ d_{n,k}]=\rho_{n}\big[\underbrace{8\pi^2S\bar{f_n^2}\delta_{\rm cn}}_{\bar{a}_{n,k}},\ \underbrace{4\pi S\mathfrak{T}\{\sigma_{tf}\}\delta_{\rm cn}}_{\bar{b}_{n,k}},\ \underbrace{8\pi^2S\bar{t_n^2}\delta_{\rm cn}}_{\bar{d}_{n,k}}\big]\label{eq:simplified SO-PD Delay_Doppler}
\end{equation}
where $S$ denotes the number of samples. 
The remaining parameters, including $\beta_{n,k}$, $\zeta_{n,k}$, $\xi_{n,k}$, $\varrho_{n,k}$, $\eta_{n,k}$, and $\kappa_{n,k}$, represent geometric spreading factors, given respectively by
\begin{subequations}
\small
\begin{align}
   [\beta_{n,k},\zeta_{n,k}]^{\rm T} =\ & \dfrac{|\alpha_{n,k}|}{c}\left(\dfrac{{\bm l}-{\bm l}_n}{\|{\bm l}_n-{\bm l}\|_2}+\dfrac{{\bm l}-{\bm l}_k}{\|{\bm l}_k-{\bm l}\|_2}\right),\\
   [\xi_{n,k},\varrho_{n,k}]^{\rm T}=\ & \dfrac{f_c|\alpha_{n,k}|}{c}\left(\dfrac{{\bm l}_n-{\bm l}}{\|{\bm l}_n-{\bm l}\|_2}+\dfrac{{\bm l}_k-{\bm l}}{\|{\bm l}_k-{\bm l}\|_2}\right),\\
   [\eta_{n,k},\kappa_{n,k}]^{\rm T}=\ &\dfrac{f_c|\alpha_{n,k}|}{c}\bigg(({\bm l}_n-{\bm l})\dfrac{{\bm v}^{\rm T}({\bm l}_n-{\bm l})}{\|{\bm l}_n-{\bm l}\|_2^3}-\dfrac{{\bm v}}{\|{\bm l}_n-{\bm l}\|_2}\notag\\
   &+({\bm l}_k-{\bm l})\dfrac{{\bm v}^{\rm T}({\bm l}_k-{\bm l})}{\|{\bm l}_k-{\bm l}\|_2^3}-\dfrac{{\bm v}}{\|{\bm l}_k-{\bm l}\|_2}\bigg).
\end{align}
\end{subequations}

\vspace{-0.5em}
\section{Proof of \textbf{Lemma} \textbf{1}}\label{sec: proof of lemma 1}
It is obvious that $\|\Delta {\bm \rho}_{i+1}\|\rightarrow 0$ if and only if either ${\bm d}_{i}\rightarrow {\bm 0}$ or $v_i\rightarrow 0$ when $i>I$. On the other hand, in light of \eqref{eq:SO approx}, we have
\begin{equation}
\small
\Delta L({\bm \rho}_{i+1}) \approx \upsilon_i{\bm d}_{i}^{\rm T}\nabla L({\bm \rho}_{i})+\upsilon_i^2/2{\bm d}_{i}^{\rm T}\nabla^2L({\bm \rho}_{i}){\bm d}_{i}. \label{eq:SO approx L}
\end{equation} 
If ${\bm d}_{i}\rightarrow {\bm 0}$ or $v_i\rightarrow 0$ for $i>I$, we readily have $\Delta L({\bm \rho}_{i+1})\rightarrow 0$. Assume there exists $i>I$, where $\Delta L({\bm \rho}_{i+1})\rightarrow 0$ with neither ${\bm d}_{i}\rightarrow {\bm 0}$ nor $v_i\rightarrow 0$. We can solve the step size for $\Delta L({\bm \rho}_{i+1})=0$ as 
\begin{equation} \small
\upsilon_i=\dfrac{-2{\bm d}_{i}^{\rm T}\nabla L({\bm \rho}_{i})}{{\bm d}_{i}^{\rm T}\nabla^2L({\bm \rho}_{i}){\bm d}_{i}}, 
\end{equation}
which corresponds to the step size bound in the ILS with $\epsilon_{L}=0$. However, we require $\epsilon_{L}>0$ to perform ILS effectively. Thus, ${\bm d}_{i}\rightarrow {\bm 0}$ or $v_i\rightarrow 0$ are necessary for $\Delta L({\bm \rho}_{i+1})\rightarrow 0$.  \hfill$\blacksquare$

\vspace{-0.5em}
\section{Proof of \textbf{Theorem} \textbf{1}}\label{sec: proof of theorem 1}

From \eqref{eq:SO approx}, \eqref{eq:SO approx L}, and \eqref{eq:Armijo rule}, the ILS condition yields
\begin{equation} 
\small
    \begin{aligned} 
   &\upsilon_i{\bm d}_{i}^{\rm T}\nabla L({\bm \rho}_{i})+\upsilon_i^2/2{\bm d}_{i}^{\rm T}\nabla^2L({\bm \rho}_{i}){\bm d}_{i}\leq \epsilon_{L} \upsilon_i{\bm d}_{i}^{\rm T}\nabla L({\bm \rho}_{i})\notag\\
   &\Longrightarrow\ \upsilon_i\leq\dfrac{2(\epsilon_{L}-1){\bm d}_{i}^{\rm T}\nabla L({\bm \rho}_{i})}{{\bm d}_{i}^{\rm T}\nabla^2L({\bm \rho}_{i}){\bm d}_{i}}. 
   \end{aligned}
   \label{eq:step size threshold}
\end{equation} 
where positive ${\bm d}_{i}^{\rm T}\nabla^2L({\bm \rho}_{i}){\bm d}_{i}$ is considered.
With the Lipschitz gradient continuity of $L({\bm \rho})$, $\|\nabla^2L({\bm \rho}_{i})\|$ is upper bounded. This indicates a positive and finite ${\bm d}_{i}^{\rm T}\nabla^2L({\bm \rho}_{i}){\bm d}_{i}$ and thus nonzero and bounded $\upsilon_i$. Furthermore, in light of \eqref{eq:Armijo rule}, we have
\begin{equation} \small
   \Delta^2 L({\bm \rho}_{i+1}) \leq \epsilon_{L}^2 \upsilon_i^2({\bm d}_{i}^{\rm T}\nabla L({\bm \rho}_{i}))^2. 
\end{equation}
Since $L({\bm \rho})$ is monotonously decreasing and bounded, we have $\Delta^2 L({\bm \rho}_{i+1})\rightarrow 0$ when $i>I$ for a large constant $I$. This implies ${\bm d}_{i}^{\rm T}\nabla L({\bm \rho}_{i})\rightarrow 0$. However, a strict descent is satisfied with ${\bm d}_{i}^{\rm T}\nabla L({\bm \rho}_{i})<0$ at each iteration $i$ unless ${\bm d}_{i}={\bm 0}$. Consequently, sequence $\{L({\bm \rho}_{i})\}$ will converge to a stationary point with a zero descent direction ${\bm d}_{i}\rightarrow {\bm 0}$ when it has Lipschitz gradient continuity. According to \textbf{Lemma} \textbf{1}, we further have $\|\Delta {\bm \rho}_{i+1}\|\rightarrow 0$ when ${\bm d}_{i}\rightarrow {\bm 0}$. \hfill$\blacksquare$
\IEEEpeerreviewmaketitle

\small
\bibliographystyle{IEEEtran}
\bibliography{mybibfile_ISAC.bib}

\end{document}